\documentclass[11pt]{article}

\usepackage[preprint]{acl}

\usepackage{algorithm}
\usepackage{algpseudocode}
\usepackage{booktabs}
\usepackage{multirow}
\usepackage{tabularx}
\usepackage{array}
\usepackage[table]{xcolor}
\usepackage[normalem]{ulem}
\usepackage{times}
\usepackage{amsmath,amssymb}
\usepackage{latexsym}
\usepackage[most]{tcolorbox}
\usepackage{xcolor}
\usepackage{ragged2e}
\definecolor{promptbg}{RGB}{238,241,249}
\definecolor{prompttitlebg}{RGB}{220,235,255}
\definecolor{promptblue}{RGB}{30,70,255}
\usepackage{graphicx}
\usepackage{amsmath}

\newtcolorbox{promptbox}[1]{
  enhanced,
  width=0.96\columnwidth,
  colback=promptbg,
  colframe=black,
  boxrule=0.8pt,
  arc=3pt,
  left=7pt,
  right=7pt,
  top=17pt,
  bottom=7pt,
  fontupper=\footnotesize\bfseries\color{black!65},
  before upper={
    \justifying
    \setlength{\parindent}{0pt}
    \setlength{\parskip}{5pt}
  },
  title={\makebox[0.62\columnwidth][c]{#1}},
  fonttitle=\footnotesize\bfseries\color{black},
  attach boxed title to top center={yshift=-3.2mm},
  boxed title style={
    colback=prompttitlebg,
    colframe=black,
    boxrule=0.8pt,
    arc=3pt,
    left=4pt,
    right=4pt,
    top=3pt,
    bottom=3pt
  }
}
\usepackage{xspace}
\usepackage[T1]{fontenc}
\usepackage[utf8]{inputenc}

\usepackage{microtype}

\usepackage{inconsolata}
\usepackage{booktabs}
\usepackage{enumitem}
\usepackage{multirow}
\usepackage{booktabs}
\usepackage[table]{xcolor}
\usepackage{multirow}
\usepackage{adjustbox}
\usepackage[table]{xcolor}

\usepackage{graphicx}
\usepackage{amsmath}
\usepackage[normalem]{ulem}
\title{Enhancing Bug-Exposing Test Generation via Step-Level Credit Assignment and Self-Distillation from Valuable Failures}
\newcommand{\modelname}{SCALF\xspace}

\usepackage[preprint]{acl}
\author{
  \textbf{Wenjie Gan\textsuperscript{1}},
  \textbf{Zhaoxi Zhang\textsuperscript{2}},
  \textbf{Jiahui Liang\textsuperscript{3}},
  \textbf{Weikang Li\textsuperscript{3}}\thanks{Corresponding authors.},
  \\
  \textbf{Wanyuan Wang\textsuperscript{1}}\footnotemark[1],
  \textbf{Li Zhang\textsuperscript{3}},
  \textbf{Jian Dong\textsuperscript{3}}
  \\
  {\small\textsuperscript{1}School of Computer Science and Engineering,
  Southeast University}
  \\
  {\small\textsuperscript{2}School of Computer Science, Peking University}
  \\
 {\small\textsuperscript{3}Independent Researcher}
  \\
  {\small\texttt{\{ganwenjie,wywang\}@seu.edu.cn}}
  \\
  {\small\texttt{\{liangzong1995,zhangli.jojo,jian.dong.kevin\}@gmail.com}}
  \\
  {\small\texttt{zhaoxizhang25@stu.pku.edu.cn},
  \texttt{waveikd@pku.edu.cn}}
}
\begin{document}
\maketitle

\begin{abstract}
Large Language Models  have shown strong code generation capabilities, yet their outputs often contain subtle bugs.
Automatically repairing such code requires informative tests: inputs that expose faulty behavior, paired with correct expected outputs.
Recent RL-based test generation methods apply the same outcome reward to all reasoning steps, which can discourage useful reasoning and reinforce flawed steps.
They also fail to effectively learn from trajectories that find bug-triggering inputs but predict incorrect outputs, leaving their verified bug-triggering progress insufficiently exploited.
To address these limitations, we propose \modelname, a process-supervised RL approach that combines \textbf{S}tep-level \textbf{C}redit \textbf{A}ssignment with targeted \textbf{L}earning from valuable \textbf{F}ailed trajectories.
\modelname derives step-specific advantages from process rewards, allowing each reasoning stage to receive its own learning signal. 
It further routes trajectories that expose bugs but predict incorrect outputs to on-policy self-distillation, using execution feedback to learn from their valuable bug-triggering progress.
Across three benchmarks, \modelname outperforms strong prompting and post-training baselines by up to 2.97, 5.26, and 3.53 percentage points in joint success rate, output correctness rate, and bug-triggering rate, respectively.
Our code and datasets are available at \url{https://anonymous.4open.science/r/SCALF-F4D1}.
\end{abstract}

\section{Introduction}

Large Language Models (LLMs) have demonstrated strong capabilities in code generation~\citep{codellama,wizardcoder}, yet their outputs often contain subtle bugs and logical flaws~\citep{trickcatcher}.
Bug-exposing test cases provide useful feedback for program repair by pairing inputs that trigger faulty behavior with correct expected outputs~\citep{repairagent,repair,togll,molinelli2025oracles}.
Constructing such tests manually is time-consuming and difficult to scale, motivating automated bug-exposing test generation with LLMs~\citep{intut,aster}.

Existing LLM-based bug-exposing test generation methods primarily rely on prompting or task-specific post-training.
Prompting-based methods use general-purpose LLMs to generate test cases through zero- or few-shot prompts~\cite{chu2025large}.
Among post-training methods, UTGen~\cite{utgen} fine-tunes test generators on synthetically perturbed data, while ATGen~\cite{atgen} uses adversarial reinforcement learning to dynamically generate buggy programs and train test generators with execution-based outcome rewards.

\begin{figure}[t]
    \centering
    \includegraphics[width=\columnwidth]{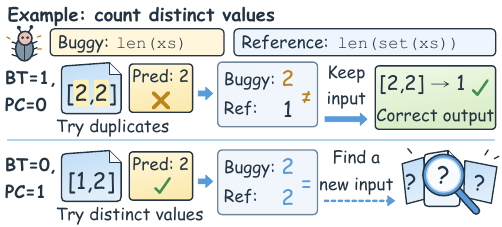}
     \caption{
     An example of different progress toward bug exposure under equal outcome rewards. BT denotes bug triggering and PC denotes prediction correctness (1: satisfied; 0: unsatisfied). For BT=1, PC=0, correcting the output completes a bug-exposing test. For BT=0, PC=1, a new input is needed to distinguish the buggy and reference programs.
     }
    \label{fig:valuable-failures}
\end{figure}

However, outcome-based optimization overlooks two key characteristics of bug-exposing test generation.
First, generating bug-triggering inputs and their expected outputs involves interdependent reasoning stages~\citep{utgen,dinella2022toga}.
Even successful trajectories may contain flawed steps, yet a uniform outcome-based learning signal rewards these steps alongside correct ones.
Although process reward models (PRMs) enable step-level supervision in mathematical~\citep{lessonPRM,zhu-etal-2025-retrieval} and code reasoning~\citep{li2025codeprm,PRM-CG}, they remain underexplored for bug-exposing test generation.
Second, partial successes offer different opportunities for further learning~\citep{togll,molinelli2025oracles,hossain2023oracle}.
As Figure~\ref{fig:valuable-failures} illustrates, a trajectory with a verified bug-triggering input but an incorrect output can yield a valid test by correcting the output.
In contrast, a trajectory with a correct output but a non-triggering input still requires discovering a new input.
However, standard GRPO does not distinguish the learning value of these partial outcomes, leaving the valuable progress in bug-triggering failures insufficiently exploited.

To address these challenges, we propose \modelname, a process-supervised RL approach that combines step-level credit assignment with targeted learning from valuable failed trajectories.
We first train a task-specific PRM to evaluate intermediate reasoning steps in bug-exposing test generation.
During RL training, \modelname derives step-specific advantages from these process rewards, allowing each reasoning stage to receive its own learning signal.
It further routes trajectories that expose bugs but predict incorrect outputs to on-policy self-distillation, using execution feedback to guide output prediction while preserving their verified bug-triggering inputs.

Across three benchmarks, \modelname-7B achieves the highest output correctness, bug-triggering, and joint success rates, exceeding the strongest baselines by up to 5.26, 3.53, and 2.97 percentage points, respectively.
In downstream program repair, its generated tests yield the highest repair success rate of 44.16\%, using seven tests and up to three repair attempts per test.

Our contributions are:
\begin{itemize}
    \item We propose \modelname for bug-exposing test generation through step-level credit assignment and targeted learning from valuable failures.

    \item We construct step-level supervision data and train a task-specific PRM for step-specific policy optimization.

   \item Experiments on three benchmarks show gains in output correctness, bug triggering, and joint success over prompting and post-training baselines, with further improvements in program repair.
\end{itemize}

\section{Related Work}

\noindent\textbf{Bug-Exposing Test Generation.}
Traditional test generation methods, including search-based testing \citep{gruber2024do}, symbolic execution \citep{busse2022combining}, and fuzzing \citep{herrera2021seed}, primarily optimize structural code coverage, yet high coverage does not necessarily translate into effective bug detection \citep{bodicoat2026understanding,10989033}.
Bug-exposing test generation instead requires both an input that triggers incorrect behavior and its correct expected output, which makes such tests directly useful for test-feedback-driven program repair \citep{wu2026debugrepair}.

\vspace{0.4\baselineskip}

\noindent\textbf{LLM-based Test Generation.}
LLM-based test generation has been explored through prompting and task-specific guidance~\citep{lemieux2023codamosa,schafer2023empirical,10.1145/3643769,togll,molinelli2025oracles}.
Post-training methods further improve bug-exposing tests: UTGen~\citep{utgen} fine-tunes on perturbed data, and ATGen~\citep{atgen} applies adversarial RL with execution-based outcome rewards.
However, ATGen assigns a single outcome reward to the entire trajectory, giving all reasoning steps the same credit.
\modelname introduces process rewards~\citep{lessonPRM,li2025codeprm,PRM-CG} so that each reasoning step receives its own learning signal, and further learns from the progress of bug-triggering failed trajectories by routing them to on-policy self-distillation~\citep{zhao2026opsd}.
\section{Problem Formulation}

Given a problem description $p$ and its corresponding incorrect solution $C_{\mathrm{buggy}}$, our goal is to generate a test case $(x,y)$, where $x$ is the input and $y$ is the expected output. Let $C_{\mathrm{ref}}$ denote a validated reference solution; a successful test case should satisfy both \textit{Prediction Correctness}, i.e., $C_{\mathrm{ref}}(x)=y$, and \textit{Bug Triggering}, i.e., $C_{\mathrm{buggy}}(x)\ne C_{\mathrm{ref}}(x)$.

We formulate the test generation as a Markov Decision Process. Following the prompt template provided in Appendix~\ref{app:prompt}, the generator $\pi_{\theta}$ produces explicit reasoning steps, mainly including \texttt{ Bug Analysis}, \texttt{ Test Design}, and \texttt{ Output Prediction}. At each step $t$, the state $s_t$ is the prefix context, and the action $a_t$ is the next structured component to generate. 
This step-wise formulation enables process rewards to provide step-specific credit for policy optimization.

\section{Dataset Construction}
\label{sec:data_construction}

To support the training of the PRM and the test generator, we construct two types of training data from \textit{CodeContests+}~\cite{codecontests_plus}, which provides a large collection of correct and incorrect solutions submitted by real users across multiple programming platforms. The construction follows a three-stage pipeline: Stage 1 filters problems for PRM training and test generator training, and Stages 2 and 3 derive step-level labels for PRM training.

\noindent\textbf{Stage 1: Difficulty Filtering.}
We first filter \textit{CodeContests+} to retain problems of moderate difficulty, since overly easy or hard problems provide weak learning signals. Specifically, for each $(\text{problem}, \text{incorrect solution})$ pair, we sample eight test-generation responses with Qwen2.5-Coder-32B-Instruct~\cite{qwen2_5_coder} and discard pairs whose responses all succeed or all fail.
For each retained problem, validated correct solutions serve as references, and clustered incorrect solutions serve as diverse buggy programs. This yields two disjoint problem sets: 2,000 problems for PRM data construction, and 1,843 problems with 15,695 $(\text{problem}, \text{incorrect solution})$ tasks for test generator training and evaluation.

% \noindent\textbf{Stage 2: MC-based Step Labeling.}
% To enable the PRM to distinguish effective processes from ineffective ones, we collect a balanced set of positive and negative test-generation trajectories. Specifically, we prompt Qwen2.5-Coder-32B-Instruct to generate multiple responses for each pair of problem description $p$ and buggy program $C_{\mathrm{buggy}}$. We execute the generated test case from each response and label the response as positive if it both exposes the bug and predicts the correct expected output; otherwise, it is labeled as negative. For each  (problem, incorrect solution) pair, we retain up to four positive responses and the same number of negative responses to reduce label-distribution bias.

\noindent\textbf{Stage 2: MC-based Step Labeling.}
To enable the PRM to distinguish effective processes from ineffective ones, we collect a balanced set of positive and negative test-generation trajectories. Specifically, we prompt Qwen2.5-Coder-32B-Instruct to generate multiple trajectories for each pair of problem description $p$ and buggy program $C_{\mathrm{buggy}}$. We execute the generated test case from each trajectory and label the trajectory as positive if it both exposes the bug and predicts the correct expected output; otherwise, it is labeled as negative. For each $(p, C_{\mathrm{buggy}})$ pair, we retain up to four positive trajectories and the same number of negative trajectories to reduce label-distribution bias.

% We then decompose each trajectory into structured reasoning steps and  derive step-level labels through Monte Carlo estimation. Given an intermediate step with prefix state $s_t$ and action $a_t$, we sample $K=8$ continuations from the updated state $s_{t+1}=\mathrm{Cat}(s_t,a_t)$ , yielding completed responses $\{\hat{r}_i\}_{i=1}^{K}$. We compute its MC score as the average execution success rate:
% {
% \begin{equation}
% MC(s_t,a_t)=MC(s_{t+1})
% =\frac{1}{K}\sum_{i=1}^{K}E(\hat{r}_i),
% \label{eq:mc_score}
% \end{equation}
% }
% where $E(\cdot)$ returns 1 if the completed response produces a bug-exposing test case with the correct expected output, and 0 otherwise.
We then decompose each trajectory into structured reasoning steps and derive step-level labels through Monte Carlo estimation. Given an intermediate step with prefix state $s_t$ and action $a_t$, we sample $K=8$ continuations from the updated state $s_{t+1}=\mathrm{Cat}(s_t,a_t)$, yielding completed trajectories $\{\hat{\tau}_i\}_{i=1}^{K}$. We compute its MC score as the average execution success rate:
{
\begin{equation}
MC(s_t,a_t)=MC(s_{t+1})
=\frac{1}{K}\sum_{i=1}^{K}E(\hat{\tau}_i),
\label{eq:mc_score}
\end{equation}
}
where $E(\cdot)$ returns 1 if the completed trajectory produces a bug-exposing test case with the correct expected output, and 0 otherwise.

We convert the MC score into a hard label, as hard labels have been shown to be more effective than soft labels for PRM training \citep{lessonPRM,scope}. Specifically, following the relative progress criterion in GenPRM \citep{GenPRM}, we define the hard label for action $a_t$ as:
% \[
% z_t =
% \begin{cases}
% 1, & \frac{MC(s_t,a_t)}{MC(s_t)} \geq \epsilon, \\
% 0, & \mathrm{otherwise}.
% \end{cases}
% \]

\begin{equation}
z_t =
\begin{cases}
1, & \frac{MC(s_t,a_t)}{MC(s_t)} \geq \epsilon, \\
0, & \mathrm{otherwise}.
\end{cases}
\label{eq:mc_label}
\end{equation}
Thus, a step is labeled as positive only when it provides sufficient relative progress toward a successful bug-exposing test case. 
We set $\epsilon=0.9$; an analysis of this threshold is
provided in Appendix~\ref{app:epsilon_analysis}.

\noindent\textbf{Stage 3: LLM-based Consistency Filtering.}
MC estimation relies on sampled continuations to infer step quality, making the resulting labels sensitive to the completion model and potentially noisy~\citep{lessonPRM}. To improve label reliability, we further introduce an LLM-based judge as an independent verifier. Specifically, we use DeepSeek-V3.2~\cite{deepseekv3.2} to assess whether each step is valid under the current context and output a binary decision. We retain only trajectories with full step-level agreement between MC labels and LLM judgments, so the final PRM training data is supported by both execution-based evidence and semantic inspection. Detailed statistics of the constructed PRM labels are provided in Appendix~\ref{app:prm_data_quality}.

Using the constructed step-level supervision data, we train a task-specific PRM to score intermediate reasoning steps. Details of PRM training are provided in Appendix~\ref{app:prm_training}.

\section{Method}
\label{sec:method}

\begin{figure*}[t]
    \centering
    \includegraphics[width=\textwidth]{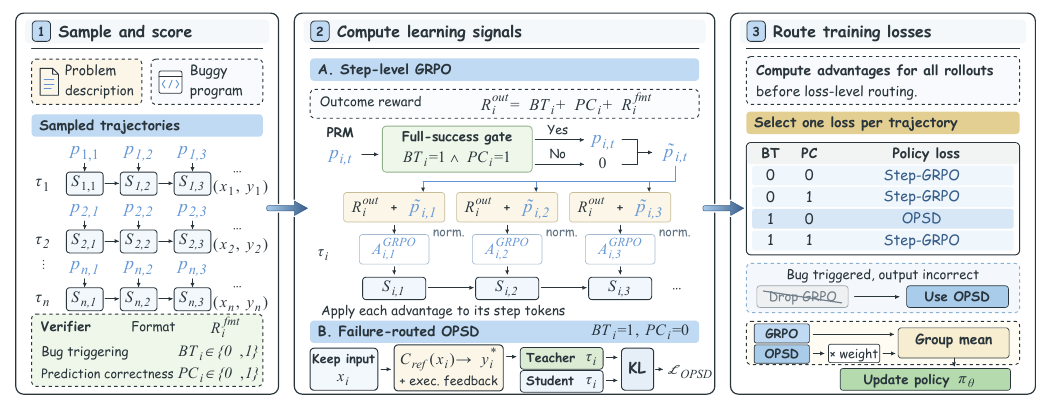}
    \caption{Overall framework of \modelname policy optimization, consisting of step-level GRPO, failure-routed OPSD, and joint loss aggregation.}
    \label{fig:sclf_framework}
\end{figure*}

This section presents the policy optimization framework of \modelname.
As illustrated in Figure~\ref{fig:sclf_framework}, \modelname combines step-level GRPO for fine-grained credit assignment with failure-routed OPSD for learning from valuable failed trajectories.
We describe the two components and their joint training objective below.

\subsection{Step-level GRPO}
\label{sec:step_grpo}

Outcome-based optimization evaluates the final test but does not directly distinguish the quality of individual reasoning steps~\citep{atgen}.
To address this limitation, we incorporate process rewards into GRPO to provide step-specific learning signals~\citep{wang-etal-2024-math,shao2024deepseekmath}.
Given $c=(p,C_{\mathrm{buggy}})$, we sample $G$ trajectories $\{\tau_i\}_{i=1}^{G}$ from $\pi_{\theta_{\mathrm{old}}}$, with $S_{i,t}$ denoting the token indices of reasoning step $t$.

Let $BT_i$ and $PC_i$ indicate whether the generated test satisfies bug triggering and prediction correctness, respectively.
The outcome reward is
\begin{equation}
R_i^{\mathrm{out}}=BT_i+PC_i+R_i^{\mathrm{fmt}},
\end{equation}
where $R_i^{\mathrm{fmt}}=-1$ for invalid formats and $0$ otherwise.
To mitigate reward hacking arising from PRM behavior on out-of-distribution trajectories~\citep{CRA}, we add the step score $p_{i,t}$ only when both objectives are satisfied:
\begin{equation}
\widetilde{R}_{i,t}
=R_i^{\mathrm{out}}
+\mathbb{I}[BT_i=1\land PC_i=1]\,p_{i,t}.
\end{equation}
This condition anchors process supervision to verified task success, allowing the PRM to distinguish reasoning quality within successful trajectories.
Further analysis is provided in Appendix~\ref{app:process_reward_activation}.
We normalize these rewards across all $G$ trajectories at each step:
\begin{equation}
A_{i,t}
=\frac{\widetilde{R}_{i,t}-\mu_t}{\sigma_t+\delta},
\end{equation}
where $\mu_t$ and $\sigma_t$ are the step-specific group mean and standard deviation, and $\delta>0$ ensures numerical stability.

Each advantage applies only to tokens in its corresponding step.
For token $u_{i,k}$ in $\tau_i$, define
\begin{equation}
\rho_{i,k}(\theta)
=\frac{\pi_{\theta}(u_{i,k}\mid u_{i,<k},c)}
{\pi_{\theta_{\mathrm{old}}}(u_{i,k}\mid u_{i,<k},c)}.
\end{equation}
Aggregating the step-specific updates yields the policy loss for trajectory $\tau_i$:
\begin{align}
&\mathcal{L}_{\mathrm{step\text{-}GRPO},i}
=\frac{1}{\sum_t|S_{i,t}|}
\sum_{t=1}^{T}\sum_{k\in S_{i,t}}
\Big[-\min\Big(
\notag\\
&\rho_{i,k}A_{i,t},\,
\operatorname{clip}(\rho_{i,k},1-\epsilon_c,1+\epsilon_c)
A_{i,t}\Big)
\\
&+\beta D_{\mathrm{KL}}(\pi_{\theta}\|\pi_{\mathrm{ref}})
\Big].
\notag
\end{align}
where $\epsilon_c$ is the clipping parameter and $\beta$ controls KL regularization.
Each reasoning step is thus optimized using its own advantage, enabling step-specific credit assignment.

\subsection{Failure-routed OPSD}
\label{sec:failure_opsd}

Correctly predicting the output for an easy input that does not trigger the bug offers limited progress toward bug exposure.
In contrast, trajectories that trigger bugs but predict incorrect outputs have already discovered a useful input; correcting the output can turn this progress into a valid bug-exposing test.
Standard GRPO does not explicitly exploit this difference in learning value.
To better learn from these valuable failures, we route them to on-policy self-distillation (OPSD)~\citep{zhao2026opsd}, replacing their step-level GRPO loss.
The routing indicator is
\begin{equation}
z_i=\mathbb{I}[BT_i=1\land PC_i=0].
\end{equation}

For each routed trajectory, we retain its input $x_i$ and obtain the correct output $y_i^{*}=C_{\mathrm{ref}}(x_i)$ through reference execution.
The verified pair $(x_i,y_i^{*})$ and execution feedback form privileged information $f_i$, enabling the self-teacher to guide output correction for the discovered bug-triggering input.

The student and self-teacher share model parameters but receive different contexts.
At each on-policy prefix $u_{i,<k}$, the student distribution $\pi^{\mathrm{S}}_{i,k}$ conditions on $(c,u_{i,<k})$, while the teacher distribution $\pi^{\mathrm{T}}_{i,k}$ additionally conditions on $f_i$.
We minimize their token-wise divergence:
\begin{equation}
\mathcal{L}_{\mathrm{OPSD},i}
=
\frac{1}{|\tau_i|}
\sum_{k=1}^{|\tau_i|}
D_{\mathrm{KL}}\!\left(
\pi^{\mathrm{S}}_{i,k}
\,\|\,\pi^{\mathrm{T}}_{i,k}
\right).
\label{eq:opsd}
\end{equation}

This provides targeted learning signals for valuable failed trajectories, using verified execution feedback to guide output prediction while retaining their bug-triggering inputs.

\subsection{Joint Training Objective}
\label{sec:joint_objective}

\modelname combines step-specific credit assignment with targeted learning from valuable failed trajectories through loss-level routing.
For each rollout group, we first compute step-level GRPO advantages across all $G$ trajectories to retain the full group-relative comparison.
We then replace the step-level GRPO loss with the OPSD loss for trajectories satisfying $z_i=1$, while keeping the GRPO loss for the remaining trajectories:
\begin{equation}
\begin{aligned}
\mathcal{L}
=\frac{1}{G}\sum_{i=1}^{G}\Big[
&(1-z_i)\mathcal{L}_{\mathrm{step\text{-}GRPO},i}\\
&+\lambda_{\mathrm{OPSD}}z_i
\mathcal{L}_{\mathrm{OPSD},i}\Big],
\end{aligned}
\end{equation}
where $z_i$ is the routing indicator defined in Section~\ref{sec:failure_opsd}, and $\lambda_{\mathrm{OPSD}}$ weights the distillation loss.
Thus, valuable failed trajectories contribute to the group-relative baseline while receiving targeted distillation updates.

\section{Experimental Setup}

\paragraph{Benchmarks.}
We evaluate on three benchmarks: \textit{CodeContests+}~\cite{codecontests_plus}, \textit{TestCase-Eval}~\cite{testcaseeval}, and \textit{MBPP+Fix (Hard)}~\cite{utgen}.
For \textit{CodeContests+}, we split the task pool at the problem level, yielding 437 evaluation tasks with no problem overlap with the training split.
For \textit{TestCase-Eval}, we retain 3,341 tasks after excluding those whose task contexts exceed 2,048 tokens.
For \textit{MBPP+Fix (Hard)}, we use the 170 evaluation tasks introduced by UTGen.
Each task provides a problem description and an incorrect solution, requiring a bug-triggering input and its correct expected output.

\paragraph{Baselines.}
We compare \modelname against two categories of methods:
\begin{itemize}[leftmargin=1.5em]
    \item \textbf{Prompting-based Methods:}
    We evaluate models from the Qwen2.5-Coder~\cite{qwen2_5_coder}, Qwen3~\cite{qwen3}, Yi-Coder~\cite{yi_coder}, and SeedCoder~\cite{seedcoder} families, directly prompting them to generate bug-exposing tests without task-specific post-training.

    \item \textbf{Post-training Methods:}
    We evaluate the 3B and 7B variants of UTGen~\cite{utgen} and ATGen~\cite{atgen}, which use supervised fine-tuning and adversarial reinforcement learning for test generation, respectively.
\end{itemize}

\paragraph{Evaluation Metrics.}
For generated tests $(x,y)$, we report three metrics :
\begin{itemize}
    \item \textbf{Prediction Correctness Rate (PCR, \%):}
    The percentage of generated test cases $(x,y)$ whose predicted output is correct, i.e., $C_{\mathrm{ref}}(x)=y$.

    \item \textbf{Bug Triggering Rate (BTR, \%):}
    The percentage of generated inputs that expose the buggy behavior, i.e., $C_{\mathrm{buggy}}(x)\neq C_{\mathrm{ref}}(x)$.

    \item \textbf{Joint Success Rate (PCR$\,\cap\,$BTR, \%):}
    The percentage of generated test cases that satisfy both prediction correctness and bug triggering.
\end{itemize}

\paragraph{Implementation Details.}
\modelname uses the 3B and 7B variants of Qwen2.5-Coder-Instruct~\cite{qwen2_5_coder}, with policy training implemented in veRL~\cite{verl}.
The SFT and RL-based baselines use the same training split as \modelname.
The PRM is initialized from Qwen2.5-Coder-7B-Instruct and trained with OpenRLHF~\cite{openrlhf}.
Training is performed on 8 NVIDIA H100 GPUs.
Detailed training settings and baseline configurations are provided in Appendix~\ref{app:implementation_details}.

\section{Evaluation Results}
\newcolumntype{C}{%
    >{\centering\arraybackslash
      \hsize=.70\hsize
      \linewidth=\hsize}X%
}

\newcolumntype{W}{%
    >{\centering\arraybackslash
      \hsize=1.60\hsize
      \linewidth=\hsize}X%
}

\begin{table*}[t]
\centering
\small
\renewcommand{\arraystretch}{1.08}
\setlength{\tabcolsep}{3.5pt}

\setlength{\aboverulesep}{1.6pt}
\setlength{\belowrulesep}{1.6pt}
\setlength{\heavyrulewidth}{0.6pt}
\setlength{\lightrulewidth}{0.45pt}
\setlength{\cmidrulewidth}{0.45pt}

\begin{tabularx}{\textwidth}{l|CCW|CCW|CCW}

\toprule

\multirow{2}{*}{\textbf{Method}}
& \multicolumn{3}{c|}{\textbf{CodeContests+}}
& \multicolumn{3}{c|}{\textbf{TestCase-Eval}}
& \multicolumn{3}{c}{\textbf{MBPP+Fix (Hard)}} \\

\cmidrule(lr){2-4}
\cmidrule(lr){5-7}
\cmidrule(lr){8-10}

& \textbf{PCR}
& \textbf{BTR}
& \textbf{PCR $\cap$ BTR}
& \textbf{PCR}
& \textbf{BTR}
& \textbf{PCR $\cap$ BTR}
& \textbf{PCR}
& \textbf{BTR}
& \textbf{PCR $\cap$ BTR} \\

\midrule
\multicolumn{10}{l}{\textbf{Baselines}} \\
\midrule

QWC-3B
& 38.67\% & 22.20\% & 8.70\%
& 31.61\% & 9.70\% & 3.47\%
& 25.88\% & 18.82\% & 12.35\% \\

QWC-7B
& 39.82\% & 22.88\% & 10.98\%
& 25.26\% & 10.30\% & 4.64\%
& 34.12\% & 24.12\% & 18.82\% \\

QWC-14B
& 45.08\% & 24.71\% & 13.50\%
& 35.17\% & 13.89\% & 6.41\%
& 35.29\% & 26.47\% & 20.59\% \\

QWC-32B
& 37.99\% & 25.40\% & 15.79\%
& 35.95\% & 14.10\% & 6.76\%
& 36.47\% & 28.82\% & 23.53\% \\

Qwen3-4B
& 49.89\% & 28.38\% & 19.68\%
& 32.45\% & 10.75\% & 3.89\%
& 28.82\% & 20.00\% & 14.71\% \\

Qwen3-8B
& 59.04\% & 28.15\% & 20.82\%
& 39.72\% & 14.61\% & 6.44\%
& 37.06\% & 26.47\% & 21.18\% \\

Qwen3-14B
& 54.46\% & 32.04\% & 20.37\%
& 43.58\% & 16.79\% & 7.60\%
& 40.00\% & 32.35\% & 26.47\% \\

Qwen3-32B
& 55.15\% & 33.18\% & 23.34\%
& 40.20\% & 18.71\% & 9.10\%
& 41.18\% & 34.12\% & \uline{28.82\%} \\

Yi-Coder
& 25.17\% & 21.05\% & 7.32\%
& 12.90\% & 7.21\% & 2.45\%
& 24.12\% & 22.94\% & 11.18\% \\

SeedCoder
& 40.27\% & 23.80\% & 12.59\%
& 25.65\% & 11.49\% & 4.67\%
& 33.53\% & 24.71\% & 15.88\% \\

UTGen-3B
& 37.76\% & 21.97\% & 9.84\%
& 27.12\% & 9.61\% & 3.71\%
& 32.35\% & 22.35\% & 15.29\% \\

UTGen-7B
& 38.44\% & 23.11\% & 10.98\%
& 28.64\% & 12.30\% & 4.91\%
& 37.06\% & 24.71\% & 20.59\% \\

ATGen-3B
& 69.11\% & 35.01\% & 20.37\%
& 60.16\% & 20.59\% & 8.98\%
& 68.82\% & 33.53\% & 25.88\% \\

ATGen-7B
& \uline{82.15\%} & \uline{38.44\%} & \uline{25.86\%}
& \uline{63.84\%} & 22.81\% & 9.22\%
& \uline{74.12\%} & \uline{37.06\%} & 28.24\% \\

\midrule
\multicolumn{10}{l}{\textbf{Ours}} \\
\midrule

\rowcolor{gray!8}
\modelname-3B
& 73.91\% & 37.76\% & 24.71\%
& 62.68\% & \uline{23.89\%} & \uline{11.10\%}
& \uline{74.12\%} & 36.47\% & 27.06\% \\

\rowcolor{gray!8}
\modelname-7B
& \textbf{87.41\%} & \textbf{41.19\%} & \textbf{28.83\%}
& \textbf{67.08\%} & \textbf{24.48\%} & \textbf{11.37\%}
& \textbf{78.82\%} & \textbf{40.59\%} & \textbf{31.18\%} \\

\bottomrule
\end{tabularx}

\caption{
Main results of \modelname and baselines.
PCR denotes Prediction Correctness Rate, BTR denotes Bug Triggering Rate,
and PCR $\cap$ BTR denotes the rate satisfying both criteria.
Best results are bolded and second-best results are underlined.
}
\label{tab:main_results}

\end{table*}
\begin{table*}[t]
\centering
\scriptsize
\setlength{\tabcolsep}{3.2pt}
\renewcommand{\arraystretch}{1.03}

\resizebox{\textwidth}{!}{%
\begin{tabular}{l|ccc|ccc|ccc}
\toprule
\multirow{2}{*}{\textbf{Variant}}
& \multicolumn{3}{c|}{\textbf{CodeContests+}}
& \multicolumn{3}{c|}{\textbf{TestCase-Eval}}
& \multicolumn{3}{c}{\textbf{MBPP+Fix (Hard)}} \\
\cmidrule(lr){2-4}
\cmidrule(lr){5-7}
\cmidrule(lr){8-10}

& \textbf{PCR}
& \textbf{BTR}
& \textbf{PCR$\,\cap\,$BTR}
& \textbf{PCR}
& \textbf{BTR}
& \textbf{PCR$\,\cap\,$BTR}
& \textbf{PCR}
& \textbf{BTR}
& \textbf{PCR$\,\cap\,$BTR} \\
\midrule

\multicolumn{10}{l}{\textbf{QWC-3B}} \\
\midrule

Standard GRPO
& 67.51\% & 35.01\% & 20.14\%
& 61.06\% & 20.02\% & 8.74\%
& 67.06\% & 29.41\% & 22.35\% \\

+ Step-GRPO
& 69.34\% & 35.24\% & 22.20\%
& 61.54\% & 21.19\% & 9.82\%
& 69.41\% & 32.35\% & 24.12\% \\

+ Routed OPSD
& 70.25\% & 36.84\% & 23.11\%
& 61.72\% & 20.98\% & 10.48\%
& 72.94\% & 35.29\% & 26.47\% \\

\rowcolor{gray!10}
\modelname
& 73.91\% & 37.76\% & 24.71\%
& 62.68\% & 23.89\% & 11.10\%
& 74.12\% & 36.47\% & 27.06\% \\

\midrule

\multicolumn{10}{l}{\textbf{QWC-7B}} \\
\midrule

Standard GRPO
& 81.69\% & 37.30\% & 25.17\%
& 62.32\% & 22.18\% & 8.98\%
& 71.18\% & 37.65\% & 27.06\% \\

+ Step-GRPO
& 84.21\% & 38.67\% & 27.69\%
& 65.25\% & 23.17\% & 9.67\%
& 75.88\% & 39.41\% & 28.24\% \\

+ Routed OPSD
& 86.73\% & 40.27\% & 28.38\%
& 64.20\% & 24.12\% & 10.84\%
& 77.06\% & 40.00\% & 30.59\% \\

\rowcolor{gray!10}
\modelname
& 87.41\% & 41.19\% & 28.83\%
& 67.08\% & 24.48\% & 11.37\%
& 78.82\% & 40.59\% & 31.18\% \\

\bottomrule
\end{tabular}%
}

\caption{Ablation results of \modelname under different training variants.}
\label{tab:ablation}
\end{table*}
\subsection{Main Results}

Table~\ref{tab:main_results} reports the main evaluation results on CodeContests+, TestCase-Eval, and MBPP+Fix (Hard).
We compare \modelname with two categories of baselines: prompting-based models, including the Qwen2.5-Coder series (abbreviated as QWC in the table), Qwen3 series, Yi-Coder-9B-Chat (abbreviated as Yi-Coder), and Seed-Coder-8B-Instruct (abbreviated as SeedCoder), and task-specific post-training baselines, including UTGen and ATGen.
Based on the experimental results, we make the following observations:

\textbf{(1) \modelname consistently improves both prediction correctness and bug triggering.}
\modelname-7B achieves the best PCR, BTR, and PCR$\,\cap\,$BTR on all three benchmarks.
On CodeContests+, it reaches 87.41\% PCR, 41.19\% BTR, and 28.83\% PCR$\,\cap\,$BTR, improving over ATGen-7B by 5.26, 2.75, and 2.97 percentage points, respectively.
Similar gains are observed on TestCase-Eval, where PCR$\,\cap\,$BTR improves from 9.22\% to 11.37\%.
On MBPP+Fix (Hard), \modelname-7B achieves 31.18\% PCR$\,\cap\,$BTR, outperforming the strongest baseline result of 28.82\%.
These results demonstrate that \modelname consistently improves both prediction correctness and bug triggering, leading to higher PCR$\,\cap\,$BTR across all three benchmarks.

\textbf{(2) \modelname provides consistent gains across backbone scales while maintaining strong parameter efficiency.}
Compared with the same-scale ATGen baselines, \modelname improves PCR$\,\cap\,$BTR at both 3B and 7B scales across all three benchmarks.
For example, on CodeContests+, \modelname improves PCR$\,\cap\,$BTR from 20.37\% to 24.71\% at 3B and from 25.86\% to 28.83\% at 7B.
Notably, despite using roughly $10\times$ fewer parameters, \modelname-3B surpasses Qwen3-32B in both PCR and BTR on all three benchmarks, and also achieves higher PCR$\,\cap\,$BTR on CodeContests+ and TestCase-Eval.
This demonstrates that the improvements of \modelname are consistent across model scales rather than being driven solely by a larger backbone.

\subsection{Ablation Studies}

To better understand the contribution of each component in \modelname, we conduct ablation studies on step-level GRPO and failure-routed OPSD.
Starting from standard GRPO, we separately introduce step-level GRPO and failure-routed OPSD, and further combine them to form the full \modelname.
We perform the same comparison on both 3B and 7B backbones.
Table~\ref{tab:ablation} reports the results, leading to the following observations.

\textbf{Step-level GRPO improves optimization effectiveness through more precise credit assignment across reasoning stages.}
Introducing step-specific credit assignment improves PCR$\,\cap\,$BTR across both model scales and all three benchmarks.
For example, on CodeContests+, Step-level GRPO improves PCR$\,\cap\,$BTR from 20.14\% to 22.20\% for the 3B model and from 25.17\% to 27.69\% for the 7B model.
Similar gains are observed on MBPP+Fix (Hard), where the 7B model improves from 27.06\% to 28.24\%.
These results show that assigning advantages at the reasoning-step level provides more effective supervision than applying a single trajectory-level signal.

\textbf{Failure-routed OPSD effectively exploits valuable partial successes and complements step-level GRPO.}
Adding failure-routed OPSD to standard GRPO consistently improves PCR$\,\cap\,$BTR across all benchmarks, reaching 23.11\%, 10.48\%, and 26.47\% for the 3B model on CodeContests+, TestCase-Eval, and MBPP+Fix (Hard), respectively.
The same trend holds for the 7B model, where PCR$\,\cap\,$BTR increases from 25.17\% to 28.38\% on CodeContests+ and from 27.06\% to 30.59\% on MBPP+Fix (Hard).
Combining failure-routed OPSD with step-level GRPO yields the strongest results, with full \modelname achieving the highest PCR$\,\cap\,$BTR across both model scales and all three benchmarks.
This demonstrates that step-level credit assignment and targeted learning from routed trajectories provide complementary benefits.

\subsection{Does Step-Level Credit Assignment Matter?}

\begin{table}[t]
\centering
\small
\setlength{\tabcolsep}{1pt}
\renewcommand{\arraystretch}{1.05}
\begin{tabularx}{\columnwidth}{@{}>{\raggedright\arraybackslash}Xccc@{}}
\toprule
\textbf{Method} & \textbf{PCR} & \textbf{BTR}
& \textbf{PCR$\,\cap\,$BTR} \\
\midrule
GRPO
& 81.69\% & 37.30\% & 25.17\% \\
PRM Aggregation (Last)
& 80.09\% & 35.01\% & 23.80\% \\
PRM Aggregation (Mean)
& 82.38\% & 36.84\% & 24.71\% \\
PRM Aggregation \mbox{(Product)}
& 83.52\% & 37.53\% & 25.86\% \\
PRM Aggregation (Min)
& \textbf{85.81\%} & 38.22\% & 26.77\% \\
Step-GRPO
& 84.21\% & \textbf{38.67\%} & \textbf{27.69\%} \\
\bottomrule
\end{tabularx}
\caption{Comparison of step-level credit assignment and trajectory-level PRM aggregation on CodeContests+ with QWC-7B.}
\label{tab:step_credit}
\end{table}

To distinguish the effect of process supervision from that of step-level credit assignment, we compare Step-GRPO with GRPO and several PRM aggregation baselines on CodeContests+.
These baselines aggregate the same step-wise PRM scores using Last, Mean, Product, or Min, and add the aggregated process reward to the outcome reward.
Group normalization then yields a single trajectory-level advantage, which is applied to all tokens in the entire trajectory.
In contrast, Step-GRPO computes a separate advantage for each reasoning step and applies it only to the corresponding tokens.

\textbf{Preserving step-specific feedback helps reduce cross-stage credit interference.}
As shown in Table~\ref{tab:step_credit}, Last and Mean aggregation yield lower joint success rates than GRPO, while Product and Min improve performance.
All four variants assign the same advantage to every reasoning step within a trajectory, so differences in step quality are not directly reflected in their respective updates.
Step-GRPO retains this distinction and achieves the highest BTR and PCR$\,\cap\,$BTR, increasing joint success from 25.17\% to 27.69\%.
Although Min aggregation achieves the highest PCR, its lower joint success rate (26.77\% vs.\ 27.69\%) shows that better output correctness does not necessarily translate into more successful bug-exposing tests.
Together, these results suggest that step-specific credit assignment helps reduce interference from shared trajectory-level updates, better supporting the combined objective of bug triggering and output correctness.

\subsection{Where Do the Gains of Failure-Routed OPSD Come From?}
\begin{figure}[t]
    \centering
    \includegraphics[width=\columnwidth]{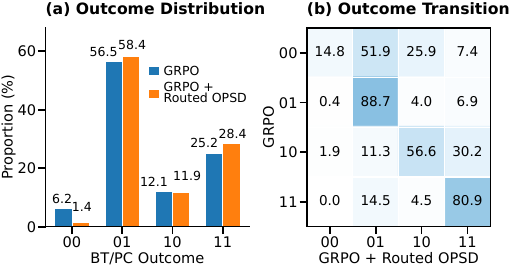}
\caption{
Outcome analysis of failure-routed OPSD on CodeContests+ with QWC-7B.
(a) Outcome distributions.
(b) Row-normalized transitions (\%) from GRPO to GRPO with routed OPSD.
}
    \label{fig:routed_analysis}
\end{figure}

% To investigate the effect of Failure-Routed OPSD, we analyze the four outcome categories in Figure~\ref{fig:routed_analysis}: $\mathrm{BT0PC0}$ (00), $\mathrm{BT0PC1}$ (01), $\mathrm{BT1PC0}$ (10), and $\mathrm{BT1PC1}$ (11). Figure~\ref{fig:routed_analysis}(a) compares the overall distribution across the four outcome categories under GRPO and GRPO with Failure-Routed OPSD. Figure~\ref{fig:routed_analysis}(b) further tracks how the same evaluation cases transition from their GRPO categories to those under GRPO with Failure-Routed OPSD.
To investigate the effect of failure-routed OPSD, we categorize test outcomes by bug triggering (BT) and prediction correctness (PC), with 1 indicating success and 0 failure.
The four categories are denoted as 00, 01, 10, and 11, with BT first and PC second.
Figure~\ref{fig:routed_analysis}(a) compares their distributions under GRPO and GRPO with failure-routed OPSD. 
Figure~\ref{fig:routed_analysis}(b) tracks category changes for the same evaluation cases, with rows indicating their outcomes under GRPO and columns indicating their outcomes under GRPO with failure-routed OPSD.
Each cell reports the percentage of cases in a given GRPO category that fall into the corresponding category after adding failure-routed OPSD.

\textbf{Failure-routed OPSD helps recover output-prediction failures while largely retaining bug-triggering success.}
As shown in Figure~\ref{fig:routed_analysis}(a), adding failure-routed OPSD increases the proportion of fully successful $\mathrm{BT1PC1}$ cases from 25.17\% to 28.38\%, while reducing $\mathrm{BT0PC0}$ from 6.18\% to 1.37\%.
Figure~\ref{fig:routed_analysis}(b) further shows that 30.2\% of cases categorized as $\mathrm{BT1PC0}$ under GRPO become fully successful with failure-routed OPSD, whereas only 13.2\% move to non-triggering categories.
The transition to full success is more frequent among $\mathrm{BT1PC0}$ cases than among $\mathrm{BT0PC1}$ cases (30.2\% vs.\ 6.9\%).
This pattern is consistent with the intended mechanism: learning from verified bug-triggering inputs helps improve output prediction while largely retaining bug-triggering success.

\subsection{Can \modelname Improve Downstream Program Repair?}
\label{sec:application}

To investigate the practical applicability of \modelname, we evaluate whether its generated tests can serve as effective feedback for downstream program repair.
We conduct this application experiment on the CodeContests+ evaluation set, where each instance consists of a problem description and an incorrect solution.

For each instance, each test generator samples $K_1$ test cases.
Each test is supplied separately to Qwen3-4B, which generates up to
$K_2$ candidate patches using that test as feedback.
An instance is considered successfully repaired if any of the
up to $K_1K_2$ candidate patches passes all official test cases.
We report the results with $K_2=3$ here, and provide the results for $K_2=1$ and $K_2=5$ in Appendix~\ref{app:repair_more_results}.

\begin{figure}[t]
    \centering
    \includegraphics[width=\columnwidth]{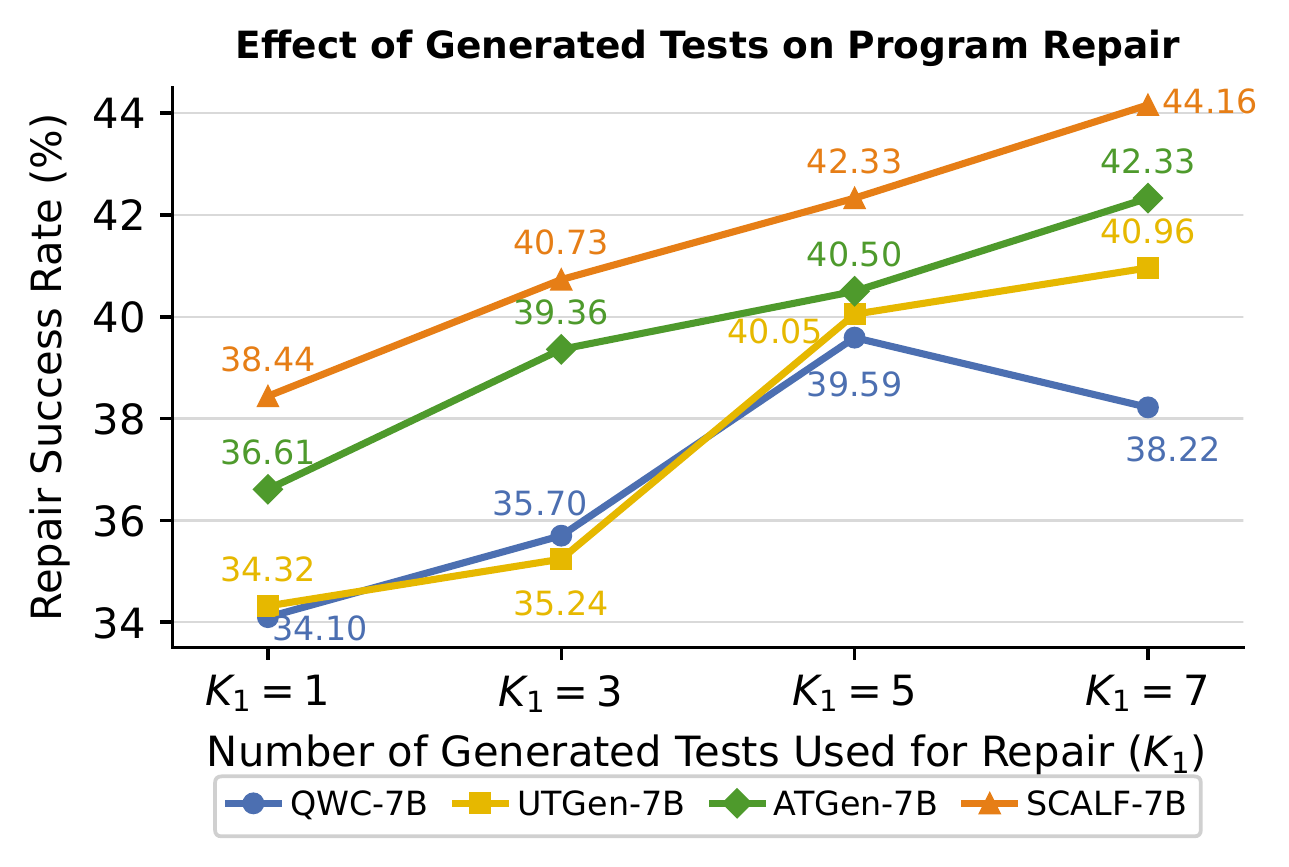}
    \caption{
    Program repair performance with different numbers of generated tests.
    We fix the maximum number of repair attempts to $K_2=3$ and vary the number of sampled generated tests, denoted by $K_1$.
    }
    \label{fig:repair_k2_3}
\end{figure}

Figure~\ref{fig:repair_k2_3} shows that \modelname-7B consistently outperforms all baselines across different values of
$K_1$.
At a small feedback budget of $K_1=1$, \modelname-7B already achieves a repair success rate of 38.44\%, outperforming the strongest baseline ATGen-7B.
As $K_1$ increases, the success rate of \modelname-7B further improves, reaching 44.16\% at $K_1=7$.
This trend suggests that \modelname-7B improves program repair by increasing the chance that the repair model observes bug-revealing evidence.
\section{Conclusion}

In this paper, we propose \modelname, a process-supervised RL approach for bug-exposing test generation.
\modelname combines step-level GRPO for step-specific credit assignment with failure-routed OPSD to learn from trajectories that trigger bugs but predict incorrect outputs.
Experiments on three benchmarks show consistent improvements over prompting-based and post-training baselines, while downstream program repair experiments demonstrate the practical value of the generated tests.
Together, these results show that learning from both intermediate reasoning steps and valuable failed trajectories improves the generation of tests that expose bugs and provide correct expected outputs.

\section*{Limitations}

Although \modelname improves bug-exposing test generation, it has several limitations.
First, to mitigate reward hacking on out-of-distribution trajectories, we apply process rewards only to fully successful trajectories, leaving failed trajectories without direct process supervision.
Even within this scope, process rewards distinguish step quality beyond identical outcome rewards, enabling finer-grained credit assignment.
Second, \modelname introduces additional training overhead through PRM scoring and self-teacher evaluation for routed trajectories.
These computations provide step-level and token-level learning signals but increase training resource requirements compared with outcome-only GRPO.

\section*{Ethical Considerations}

We discuss potential ethical considerations as follows.
(1) \textit{Intellectual property and artifact licenses.}
This work uses publicly available datasets, benchmarks, pretrained models, and software frameworks.
CodeContests+ is released under CC BY 4.0\footnote{\url{https://creativecommons.org/licenses/by/4.0/}}, and Qwen2.5-Coder models are released under Apache 2.0\footnote{\url{https://www.apache.org/licenses/LICENSE-2.0}} for most model sizes, while the 3B variant is released under the Qwen Research License\footnote{\url{https://huggingface.co/Qwen/Qwen2.5-Coder-3B-Instruct/blob/main/LICENSE}}.
The main software frameworks used in our experiments, including veRL, OpenRLHF, and vLLM, are released under Apache 2.0.
For MBPP+Fix (Hard), we follow the UTGenDebug release, whose evaluation code is distributed under the MIT license\footnote{\url{https://github.com/archiki/UTGenDebug/blob/main/LICENSE}}.
All artifacts are used for research purposes consistent with their released licenses, terms of use, and intended use when specified.
Our released code and constructed datasets are provided for research use, with redistribution limited to materials permitted by the licenses and terms of the original artifacts.
(2) \textit{Privacy and sensitive content.}
Our data primarily consists of programming problem statements, code submissions, and generated test cases, and does not require personal identifiers, demographic attributes, or user profiles.
We exclude directly identifying metadata from released artifacts and do not intentionally collect or introduce sensitive personal information.
(3) \textit{Potential risks.}
Our work targets controlled programming benchmarks for bug-exposing test generation and test-feedback-driven program repair.
Generated tests with incorrect expected outputs may provide misleading feedback to downstream repair systems.
Therefore, practical deployment, particularly in safety-critical software settings, should combine generated tests with execution-based validation, additional verification, or human review.

% Bibliography entries for the entire Anthology, followed by custom entries
%\bibliography{custom,anthology-overleaf-1,anthology-overleaf-2}

% Custom bibliography entries only
\bibliography{custom}

@inproceedings{hossain2023oracle,
  title     = {Neural-Based Test Oracle Generation:
               A Large-Scale Evaluation and Lessons Learned},
  author    = {Hossain, Soneya Binta and Filieri, Antonio
               and Dwyer, Matthew B. and Elbaum, Sebastian
               and Visser, Willem},
  booktitle = {Proceedings of the 31st ACM Joint European
               Software Engineering Conference and Symposium
               on the Foundations of Software Engineering},
  year      = {2023},
  publisher = {Association for Computing Machinery},
  doi       = {10.1145/3611643.3616265},
  pages = {120--132}
}

@article{shao2024deepseekmath,
  title   = {{DeepSeekMath}: Pushing the Limits of Mathematical
             Reasoning in Open Language Models},
  author  = {Shao, Zhihong and Wang, Peiyi and Zhu, Qihao
             and Xu, Runxin and Song, Junxiao and Bi, Xiao
             and Zhang, Haowei and Zhang, Mingchuan
             and Li, Y. K. and Wu, Y. and Guo, Daya},
  journal = {arXiv preprint arXiv:2402.03300},
  year    = {2024},
  doi     = {10.48550/arXiv.2402.03300},
  url     = {https://arxiv.org/abs/2402.03300}
}

@inproceedings{
utgen,
title={Learning to Generate Unit Tests for Automated Debugging},
author={Archiki Prasad and Elias Stengel-Eskin and Justin Chen and Zaid Khan and Mohit Bansal},
booktitle={Second Conference on Language Modeling},
year={2025},
url={https://openreview.net/forum?id=yeVBHPLXxi}
}

@inproceedings{
atgen,
title={{ATG}en: Adversarial Reinforcement Learning for Test Case Generation},
author={Qingyao Li and Xinyi Dai and Weiwen Liu and Xiangyang Li and Yasheng Wang and Ruiming Tang and Yong Yu and Weinan Zhang},
booktitle={The Fourteenth International Conference on Learning Representations},
year={2026},
url={https://openreview.net/forum?id=Sxj4o3qXtl}
}

@inproceedings{codecontests_plus,
    title = "{C}ode{C}ontests+: High-Quality Test Case Generation for Competitive Programming",
    author = "Wang, Zihan  and
      Liu, Siyao  and
      Sun, Yang  and
      Ding, Ming  and
      Li, Hongyan",
    editor = "Christodoulopoulos, Christos  and
      Chakraborty, Tanmoy  and
      Rose, Carolyn  and
      Peng, Violet",
    booktitle = "Findings of the Association for Computational Linguistics: EMNLP 2025",
    month = nov,
    year = "2025",
    address = "Suzhou, China",
    publisher = "Association for Computational Linguistics",
    url = "https://aclanthology.org/2025.findings-emnlp.299/",
    doi = "10.18653/v1/2025.findings-emnlp.299",
    pages = "5576--5600",
    ISBN = "979-8-89176-335-7"
}

@article{qwen2_5_coder,
  author       = {Binyuan Hui and
                  Jian Yang and
                  Zeyu Cui and
                  Jiaxi Yang and
                  Dayiheng Liu and
                  Lei Zhang and
                  Tianyu Liu and
                  Jiajun Zhang and
                  Bowen Yu and
                  Kai Dang and
                  An Yang and
                  Rui Men and
                  Fei Huang and
                  Xingzhang Ren and
                  Xuancheng Ren and
                  Jingren Zhou and
                  Junyang Lin},
  title        = {{Qwen2.5-Coder} Technical Report},
  journal      = {CoRR},
  volume       = {abs/2409.12186},
  year         = {2024},
  url          = {https://doi.org/10.48550/arXiv.2409.12186},
  doi          = {10.48550/ARXIV.2409.12186},
  eprinttype   = {arXiv},
  eprint       = {2409.12186},
  bibsource    = {dblp computer science bibliography, https://dblp.org}
}

@article{qwen3,
  author       =  {{Qwen Team}},
  title        = {{Qwen3} Technical Report},
  journal      = {CoRR},
  volume       = {abs/2505.09388},
  year         = {2025},
  url          = {https://doi.org/10.48550/arXiv.2505.09388},
  doi          = {10.48550/ARXIV.2505.09388},
  eprinttype   = {arXiv},
  eprint       = {2505.09388},
  bibsource    = {dblp computer science bibliography, https://dblp.org}
}

@article{GenPRM,
  author = {Jian Zhao and Runze Liu and Kaiyan Zhang and Zhimu Zhou and Junqi Gao and Dong Li and Jiafei Lyu and Zhouyi Qian and Biqing Qi and Xiu Li and Bowen Zhou},
  title = {{GenPRM}: Scaling Test-Time Compute of Process Reward Models via Generative Reasoning},
  journal = {Proceedings of the AAAI Conference on Artificial Intelligence},
  volume = {40},
  number = {41},
  pages = {34932--34940},
  year = {2026},
  month = mar,
  doi = {10.1609/aaai.v40i41.40797},
  url = {https://ojs.aaai.org/index.php/AAAI/article/view/40797}
}

@inproceedings{lessonPRM,
    title = "The Lessons of Developing Process Reward Models in Mathematical Reasoning",
    author = "Zhang, Zhenru  and
      Zheng, Chujie  and
      Wu, Yangzhen  and
      Zhang, Beichen  and
      Lin, Runji  and
      Yu, Bowen  and
      Liu, Dayiheng  and
      Zhou, Jingren  and
      Lin, Junyang",
    editor = "Che, Wanxiang  and
      Nabende, Joyce  and
      Shutova, Ekaterina  and
      Pilehvar, Mohammad Taher",
    booktitle = "Findings of the Association for Computational Linguistics: ACL 2025",
    month = jul,
    year = "2025",
    address = "Vienna, Austria",
    publisher = "Association for Computational Linguistics",
    url = "https://aclanthology.org/2025.findings-acl.547/",
    doi = "10.18653/v1/2025.findings-acl.547",
    pages = "10495--10516",
    ISBN = "979-8-89176-256-5"
}

@inproceedings{scope,
    title = "{SCOPE}: Compress Mathematical Reasoning Steps for Efficient Automated Process Annotation",
    author = "Xu, Huimin  and
      Mao, Xin  and
      Li, Feng-Lin  and
      Wu, Xiaobao  and
      Chen, Wang  and
      Zhang, Wei  and
      Luu, Anh Tuan",
    editor = "Che, Wanxiang  and
      Nabende, Joyce  and
      Shutova, Ekaterina  and
      Pilehvar, Mohammad Taher",
    booktitle = "Findings of the Association for Computational Linguistics: ACL 2025",
    month = jul,
    year = "2025",
    address = "Vienna, Austria",
    publisher = "Association for Computational Linguistics",
    url = "https://aclanthology.org/2025.findings-acl.1251/",
    doi = "10.18653/v1/2025.findings-acl.1251",
    pages = "24382--24394",
    ISBN = "979-8-89176-256-5"
}

@article{deepseekv3.2,
  author       = {{DeepSeek-AI}},
  title        = {{DeepSeek-V3.2}: Pushing the Frontier of Open Large Language Models},
  journal      = {CoRR},
  volume       = {abs/2512.02556},
  year         = {2025},
  url          = {https://doi.org/10.48550/arXiv.2512.02556},
  doi          = {10.48550/ARXIV.2512.02556},
  eprinttype   = {arXiv},
  eprint       = {2512.02556},
  bibsource    = {dblp computer science bibliography, https://dblp.org}
}

@inproceedings{PRM-CG,
    title = "Process-Supervised Reinforcement Learning for Code Generation",
    author = "Ye, Yufan  and
      Zhang, Ting  and
      Jiang, Wenbin  and
      Huang, Hua",
    editor = "Christodoulopoulos, Christos  and
      Chakraborty, Tanmoy  and
      Rose, Carolyn  and
      Peng, Violet",
    booktitle = "Proceedings of the 2025 Conference on Empirical Methods in Natural Language Processing",
    month = nov,
    year = "2025",
    address = "Suzhou, China",
    publisher = "Association for Computational Linguistics",
    url = "https://aclanthology.org/2025.emnlp-main.719/",
    doi = "10.18653/v1/2025.emnlp-main.719",
    pages = "14213--14226",
    ISBN = "979-8-89176-332-6"
}

@inproceedings{CRA,
  author       = {Ruike Song and
                  Zeen Song and
                  Huijie Guo and
                  Wenwen Qiang},
  editor       = {Sven Koenig and
                  Chad Jenkins and
                  Matthew E. Taylor},
  title        = {Causal Reward Adjustment: Mitigating Reward Hacking in External Reasoning
                  via Backdoor Correction},
  booktitle    = {Fortieth {AAAI} Conference on Artificial Intelligence, Thirty-Eighth
                  Conference on Innovative Applications of Artificial Intelligence,
                  Sixteenth Symposium on Educational Advances in Artificial Intelligence,
                  {AAAI} 2026, Singapore, January 20-27, 2026},
  pages        = {33019--33027},
  publisher    = {{AAAI} Press},
  year         = {2026},
  url          = {https://doi.org/10.1609/aaai.v40i39.40584},
  doi          = {10.1609/AAAI.V40I39.40584},
  bibsource    = {dblp computer science bibliography, https://dblp.org}
}

@inproceedings{
lightman2024lets,
title={Let's Verify Step by Step},
author={Hunter Lightman and Vineet Kosaraju and Yuri Burda and Harrison Edwards and Bowen Baker and Teddy Lee and Jan Leike and John Schulman and Ilya Sutskever and Karl Cobbe},
booktitle={The Twelfth International Conference on Learning Representations},
year={2024},
url={https://openreview.net/forum?id=v8L0pN6EOi}
}

@inproceedings{wang-etal-2024-math,
    title = {{Math-Shepherd}: Verify and Reinforce {LLM}s Step-by-step without Human Annotations},
    author = "Wang, Peiyi  and
      Li, Lei  and
      Shao, Zhihong  and
      Xu, Runxin  and
      Dai, Damai  and
      Li, Yifei  and
      Chen, Deli  and
      Wu, Yu  and
      Sui, Zhifang",
    editor = "Ku, Lun-Wei  and
      Martins, Andre  and
      Srikumar, Vivek",
    booktitle = "Proceedings of the 62nd Annual Meeting of the Association for Computational Linguistics (Volume 1: Long Papers)",
    month = aug,
    year = "2024",
    address = "Bangkok, Thailand",
    publisher = "Association for Computational Linguistics",
    url = "https://aclanthology.org/2024.acl-long.510/",
    doi = "10.18653/v1/2024.acl-long.510",
    pages = "9426--9439"
}

@inproceedings{zhu-etal-2025-retrieval,
    title = "Retrieval-Augmented Process Reward Model for Generalizable Mathematical Reasoning",
    author = "Zhu, Jiachen  and
      Zheng, Congmin  and
      Lin, Jianghao  and
      Du, Kounianhua  and
      Wen, Ying  and
      Yu, Yong  and
      Wang, Jun  and
      Zhang, Weinan",
    editor = "Che, Wanxiang  and
      Nabende, Joyce  and
      Shutova, Ekaterina  and
      Pilehvar, Mohammad Taher",
    booktitle = "Findings of the Association for Computational Linguistics: ACL 2025",
    month = jul,
    year = "2025",
    address = "Vienna, Austria",
    publisher = "Association for Computational Linguistics",
    url = "https://aclanthology.org/2025.findings-acl.444/",
    doi = "10.18653/v1/2025.findings-acl.444",
    pages = "8453--8468",
    ISBN = "979-8-89176-256-5",
}

@inproceedings{gruber2024do,
author = {Gruber, Martin and Roslan, Muhammad Firhard and Parry, Owain and Scharnb\"{o}ck, Fabian and McMinn, Phil and Fraser, Gordon},
title = {Do Automatic Test Generation Tools Generate Flaky Tests?},
year = {2024},
isbn = {9798400702174},
publisher = {Association for Computing Machinery},
address = {New York, NY, USA},
url = {https://doi.org/10.1145/3597503.3608138},
doi = {10.1145/3597503.3608138},
booktitle = {Proceedings of the IEEE/ACM 46th International Conference on Software Engineering},
articleno = {47},
numpages = {12},
location = {Lisbon, Portugal},
series = {ICSE '24},
  pages = {1--12}
}

@inproceedings{busse2022combining,
author = {Busse, Frank and Gharat, Pritam and Cadar, Cristian and Donaldson, Alastair F.},
title = {Combining static analysis error traces with dynamic symbolic execution (experience paper)},
year = {2022},
isbn = {9781450393799},
publisher = {Association for Computing Machinery},
address = {New York, NY, USA},
url = {https://doi.org/10.1145/3533767.3534384},
doi = {10.1145/3533767.3534384},
booktitle = {Proceedings of the 31st ACM SIGSOFT International Symposium on Software Testing and Analysis},
pages = {568--579},
numpages = {12},
location = {Virtual, South Korea},
series = {ISSTA 2022}
}

@inproceedings{herrera2021seed,
author = {Herrera, Adrian and Gunadi, Hendra and Magrath, Shane and Norrish, Michael and Payer, Mathias and Hosking, Antony L.},
title = {Seed selection for successful fuzzing},
year = {2021},
isbn = {9781450384599},
publisher = {Association for Computing Machinery},
address = {New York, NY, USA},
url = {https://doi.org/10.1145/3460319.3464795},
doi = {10.1145/3460319.3464795},
booktitle = {Proceedings of the 30th ACM SIGSOFT International Symposium on Software Testing and Analysis},
pages = {230--243},
numpages = {14},
location = {Virtual, Denmark},
series = {ISSTA 2021}
}

@inproceedings{bodicoat2026understanding,
  title={Understanding {LLM}-Driven Test Oracle Generation},
  author={Bodicoat, Adam and Jahangirova, Gunel and Terragni, Valerio},
  booktitle={2025 2nd IEEE/ACM International Conference on AI-powered Software (AIware)},
  pages={29--39},
  year={2025},
  organization={IEEE}
}

@article{wu2026debugrepair,
  author       = {Linhao Wu and
                  Yifei Pei and
                  Zhen Yang and
                  Kainan Li and
                  Zhonghang Lu and
                  Hao Tan and
                  Xiran Lyu and
                  Jia Li and
                  Yizhou Chen and
                  Pengyu Xue and
                  Kunwu Zheng and
                  Dan Hao},
  title        = {{DebugRepair}: Enhancing {LLM}-Based Automated Program Repair via Self-Directed Debugging},
  journal      = {CoRR},
  volume       = {abs/2604.19305},
  year         = {2026},
  url          = {https://doi.org/10.48550/arXiv.2604.19305},
  doi          = {10.48550/ARXIV.2604.19305},
  eprinttype   = {arXiv},
  eprint       = {2604.19305},
  bibsource    = {dblp computer science bibliography, https://dblp.org}
}

@inproceedings{lemieux2023codamosa,
  author       = {Caroline Lemieux and
                  Jeevana Priya Inala and
                  Shuvendu K. Lahiri and
                  Siddhartha Sen},
  title        = {{CodaMosa}: Escaping Coverage Plateaus in Test Generation with Pre-trained Large Language Models},
  booktitle    = {45th {IEEE/ACM} International Conference on Software Engineering,
                  {ICSE} 2023, Melbourne, Australia, May 14-20, 2023},
  pages        = {919--931},
  publisher    = {{IEEE}},
  year         = {2023},
  url          = {https://doi.org/10.1109/ICSE48619.2023.00085},
  doi          = {10.1109/ICSE48619.2023.00085},
  bibsource    = {dblp computer science bibliography, https://dblp.org}
}

@article{schafer2023empirical,
  title={An empirical evaluation of using large language models for automated unit test generation},
  author={Sch{\"a}fer, Max and Nadi, Sarah and Eghbali, Aryaz and Tip, Frank},
  journal={IEEE Transactions on Software Engineering},
  volume={50},
  number={1},
  pages={85--105},
  year={2024},
  publisher={IEEE}
}

@article{chu2025large,
  title={Large Language Models for Unit Test Generation: Achievements, Challenges, and Opportunities},
  author={Chu, Bei and Feng, Yang and Liu, Kui and Guo, Zhaoqiang and Zhang, Yichi and Shi, Hange and Nan, Zifan and Xu, Baowen},
  journal={arXiv preprint arXiv:2511.21382},
  year={2025}
}

@inproceedings{testcaseeval,
    title = "Can {LLM}s Generate High-Quality Test Cases for Algorithm Problems? {T}est{C}ase-Eval: A Systematic Evaluation of Fault Coverage and Exposure",
    author = "Yang, Zheyuan  and
      Kuang, Zexi  and
      Xia, Xue  and
      Zhao, Yilun",
    editor = "Che, Wanxiang  and
      Nabende, Joyce  and
      Shutova, Ekaterina  and
      Pilehvar, Mohammad Taher",
    booktitle = "Proceedings of the 63rd Annual Meeting of the Association for Computational Linguistics (Volume 2: Short Papers)",
    month = jul,
    year = "2025",
    address = "Vienna, Austria",
    publisher = "Association for Computational Linguistics",
    url = "https://aclanthology.org/2025.acl-short.82/",
    doi = "10.18653/v1/2025.acl-short.82",
    pages = "1050--1063",
    ISBN = "979-8-89176-252-7",
}

@inproceedings{verl,
author = {Sheng, Guangming and Zhang, Chi and Ye, Zilingfeng and Wu, Xibin and Zhang, Wang and Zhang, Ru and Peng, Yanghua and Lin, Haibin and Wu, Chuan},
title = {{HybridFlow}: A Flexible and Efficient {RLHF} Framework},
year = {2025},
isbn = {9798400711961},
publisher = {Association for Computing Machinery},
address = {New York, NY, USA},
url = {https://doi.org/10.1145/3689031.3696075},
doi = {10.1145/3689031.3696075},
booktitle = {Proceedings of the Twentieth European Conference on Computer Systems},
pages = {1279--1297},
numpages = {19},
location = {Rotterdam, Netherlands},
series = {EuroSys '25}
}

@inproceedings{openrlhf,
    title = "{O}pen{RLHF}: A Ray-based Easy-to-use, Scalable and High-performance {RLHF} Framework",
    author = "Hu, Jian  and
      Wu, Xibin  and
      Shen, Wei  and
      Liu, Jason Klein  and
      Wang, Weixun  and
      Jiang, Songlin  and
      Wang, Haoran  and
      Chen, Hao  and
      Chen, Bin  and
      Fang, Wenkai  and
      Xianyu  and
      Cao, Yu  and
      Xu, Haotian  and
      Liu, Yiming",
    editor = {Habernal, Ivan  and
      Schulam, Peter  and
      Tiedemann, J{\"o}rg},
    booktitle = "Proceedings of the 2025 Conference on Empirical Methods in Natural Language Processing: System Demonstrations",
    month = nov,
    year = "2025",
    address = "Suzhou, China",
    publisher = "Association for Computational Linguistics",
    url = "https://aclanthology.org/2025.emnlp-demos.48/",
    doi = "10.18653/v1/2025.emnlp-demos.48",
    pages = "656--666",
    ISBN = "979-8-89176-334-0"
}

@inproceedings{trickcatcher,
    title = "{LLM}-Powered Test Case Generation for Detecting Bugs in Plausible Programs",
    author = "Liu, Kaibo  and
      Chen, Zhenpeng  and
      Liu, Yiyang  and
      Zhang, Jie M.  and
      Harman, Mark  and
      Han, Yudong  and
      Ma, Yun  and
      Dong, Yihong  and
      Li, Ge  and
      Huang, Gang",
    editor = "Che, Wanxiang  and
      Nabende, Joyce  and
      Shutova, Ekaterina  and
      Pilehvar, Mohammad Taher",
    booktitle = "Proceedings of the 63rd Annual Meeting of the Association for Computational Linguistics (Volume 1: Long Papers)",
    month = jul,
    year = "2025",
    address = "Vienna, Austria",
    publisher = "Association for Computational Linguistics",
    url = "https://aclanthology.org/2025.acl-long.20/",
    doi = "10.18653/v1/2025.acl-long.20",
    pages = "430--440",
    ISBN = "979-8-89176-251-0"
}

@INPROCEEDINGS{intut,
  author={Nan, Zifan and Guo, Zhaoqiang and Liu, Kui and Xia, Xin},
  booktitle={2025 IEEE/ACM 47th International Conference on Software Engineering (ICSE)}, 
  title={Test Intention Guided {LLM}-Based Unit Test Generation}, 
  year={2025},
  volume={},
  number={},
  pages={1026--1038},
  doi={10.1109/ICSE55347.2025.00243}}

@INPROCEEDINGS{aster,
  author={Pan, Rangeet and Kim, Myeongsoo and Krishna, Rahul and Pavuluri, Raju and Sinha, Saurabh},
  booktitle={2025 IEEE/ACM 47th International Conference on Software Engineering: Software Engineering in Practice (ICSE-SEIP)}, 
  title={{ASTER}: Natural and Multi-Language Unit Test Generation with {LLMs}}, 
  year={2025},
  volume={},
  number={},
  pages={413--424},
  doi={10.1109/ICSE-SEIP66354.2025.00042}}

@inproceedings{togll,
author = {Hossain, Soneya Binta and Dwyer, Matthew B.},
title = {{TOGLL}: Correct and Strong Test Oracle Generation with {LLMs}},
year = {2025},
isbn = {9798331505691},
publisher = {IEEE Press},
url = {https://doi.org/10.1109/ICSE55347.2025.00098},
doi = {10.1109/ICSE55347.2025.00098},
booktitle = {Proceedings of the IEEE/ACM 47th International Conference on Software Engineering},
pages = {1475--1487},
numpages = {13},
location = {Ottawa, Ontario, Canada},
series = {ICSE '25}
}

@INPROCEEDINGS{molinelli2025oracles,
  author={Molinelli, Davide and Grazia, Luca Di and Martin-Lopez, Alberto and Ernst, Michael D. and Pezzè, Mauro},
  booktitle={2025 40th IEEE/ACM International Conference on Automated Software Engineering (ASE)}, 
  title={Do {LLMs} Generate Useful Test Oracles? An Empirical Study with an Unbiased Dataset}, 
  year={2025},
  volume={},
  number={},
  pages={278--290},
  doi={10.1109/ASE63991.2025.00031}}

@inproceedings{repairagent,
author = {Bouzenia, Islem and Devanbu, Premkumar and Pradel, Michael},
title = {{RepairAgent}: An Autonomous, {LLM}-Based Agent for Program Repair},
year = {2025},
isbn = {9798331505691},
publisher = {IEEE Press},
url = {https://doi.org/10.1109/ICSE55347.2025.00157},
doi = {10.1109/ICSE55347.2025.00157},
booktitle = {Proceedings of the IEEE/ACM 47th International Conference on Software Engineering},
pages = {2188--2200},
numpages = {13},
location = {Ottawa, Ontario, Canada},
series = {ICSE '25}
}

@inproceedings{repair,
    title = "{R}e{P}air: Automated Program Repair with Process-based Feedback",
    author = "Zhao, Yuze  and
      Huang, Zhenya  and
      Ma, Yixiao  and
      Li, Rui  and
      Zhang, Kai  and
      Jiang, Hao  and
      Liu, Qi  and
      Zhu, Linbo  and
      Su, Yu",
    editor = "Ku, Lun-Wei  and
      Martins, Andre  and
      Srikumar, Vivek",
    booktitle = "Findings of the Association for Computational Linguistics: ACL 2024",
    month = aug,
    year = "2024",
    address = "Bangkok, Thailand",
    publisher = "Association for Computational Linguistics",
    url = "https://aclanthology.org/2024.findings-acl.973/",
    doi = "10.18653/v1/2024.findings-acl.973",
    pages = "16415--16429"
}

@inproceedings{li2025codeprm,
    title = "{C}ode{PRM}: Execution Feedback-enhanced Process Reward Model for Code Generation",
    author = "Li, Qingyao  and
      Dai, Xinyi  and
      Li, Xiangyang  and
      Zhang, Weinan  and
      Wang, Yasheng  and
      Tang, Ruiming  and
      Yu, Yong",
    editor = "Che, Wanxiang  and
      Nabende, Joyce  and
      Shutova, Ekaterina  and
      Pilehvar, Mohammad Taher",
    booktitle = "Findings of the Association for Computational Linguistics: ACL 2025",
    month = jul,
    year = "2025",
    address = "Vienna, Austria",
    publisher = "Association for Computational Linguistics",
    url = "https://aclanthology.org/2025.findings-acl.428/",
    doi = "10.18653/v1/2025.findings-acl.428",
    pages = "8169--8182",
    ISBN = "979-8-89176-256-5"
}

@INPROCEEDINGS{10989033,
  author={Abdullin, Azat and Derakhshanfar, Pouria and Panichella, Annibale},
  booktitle={2025 IEEE Conference on Software Testing, Verification and Validation (ICST)}, 
  title={Test Wars: A Comparative Study of {SBST}, Symbolic Execution, and {LLM}-Based Approaches to Unit Test Generation}, 
  year={2025},
  volume={},
  number={},
  pages={221--232},
  doi={10.1109/ICST62969.2025.10989033}}

@article{10.1145/3643769,
author = {Ryan, Gabriel and Jain, Siddhartha and Shang, Mingyue and Wang, Shiqi and Ma, Xiaofei and Ramanathan, Murali Krishna and Ray, Baishakhi},
title = {Code-Aware Prompting: A Study of Coverage-Guided Test Generation in Regression Setting using {LLM}},
year = {2024},
issue_date = {July 2024},
publisher = {Association for Computing Machinery},
address = {New York, NY, USA},
volume = {1},
number = {FSE},
url = {https://doi.org/10.1145/3643769},
doi = {10.1145/3643769},
journal = {Proc. ACM Softw. Eng.},
month = jul,
articleno = {43},
numpages = {21},
  pages = {951--971}
}

@inproceedings{Posterior-GRPO,
  title     = {{ReCode}: Reinforcing Code Generation with
               Reasoning-Process Rewards},
  author    = {Fan, Lishui and Zhang, Yu and Chen, Mouxiang
               and Liu, Zhongxin},
  booktitle = {Proceedings of the 64th Annual Meeting of the
               Association for Computational Linguistics
               (Volume 1: Long Papers)},
  year      = {2026},
  month     = jul,
  publisher = {Association for Computational Linguistics},
  pages     = {43896--43914},
  doi       = {10.18653/v1/2026.acl-long.2031},
  url       = {https://aclanthology.org/2026.acl-long.2031/}
}

@article{yi_coder,
  title={{Yi}: Open Foundation Models by {01.AI}},
  author={{01.AI}},
  journal={arXiv preprint arXiv:2403.04652},
  year={2024}
}

@article{codellama,
  author       = {Baptiste Rozi{\`{e}}re and
                  Jonas Gehring and
                  Fabian Gloeckle and
                  Sten Sootla and
                  Itai Gat and
                  Xiaoqing Ellen Tan and
                  Yossi Adi and
                  Jingyu Liu and
                  Tal Remez and
                  J{\'{e}}r{\'{e}}my Rapin and
                  Artyom Kozhevnikov and
                  Ivan Evtimov and
                  Joanna Bitton and
                  Manish Bhatt and
                  Cristian Canton{-}Ferrer and
                  Aaron Grattafiori and
                  Wenhan Xiong and
                  Alexandre D{\'{e}}fossez and
                  Jade Copet and
                  Faisal Azhar and
                  Hugo Touvron and
                  Louis Martin and
                  Nicolas Usunier and
                  Thomas Scialom and
                  Gabriel Synnaeve},
  title        = {{Code Llama}: Open Foundation Models for Code},
  journal      = {CoRR},
  volume       = {abs/2308.12950},
  year         = {2023},
  url          = {https://doi.org/10.48550/arXiv.2308.12950},
  doi          = {10.48550/ARXIV.2308.12950},
  eprinttype   = {arXiv},
  eprint       = {2308.12950},
  bibsource    = {dblp computer science bibliography, https://dblp.org}
}

@inproceedings{
wizardcoder,
title={{WizardCoder}: Empowering Code Large Language Models with {Evol-Instruct}},
author={Ziyang Luo and Can Xu and Pu Zhao and Qingfeng Sun and Xiubo Geng and Wenxiang Hu and Chongyang Tao and Jing Ma and Qingwei Lin and Daxin Jiang},
booktitle={The Twelfth International Conference on Learning Representations},
year={2024},
url={https://openreview.net/forum?id=UnUwSIgK5W}
}

@article{seedcoder,
  publtype={informal},
  author={Yuyu Zhang and Jing Su and Yifan Sun and Chenguang Xi and Xia Xiao and Shen Zheng and Anxiang Zhang and Kaibo Liu and Daoguang Zan and Tao Sun and Jinhua Zhu and Shulin Xin and Dong Huang and Yetao Bai and Lixin Dong and Chao Li and Jianchong Chen and Hanzhi Zhou and Yifan Huang and Guanghan Ning and Xierui Song and Jiaze Chen and Siyao Liu and Kai Shen and Liang Xiang and Yonghui Wu},
  title={{Seed-Coder}: Let the Code Model Curate Data for Itself},
  year={2025},
  month={June},
  cdate={1748736000000},
  journal={CoRR},
  volume={abs/2506.03524},
  url={https://doi.org/10.48550/arXiv.2506.03524}
}

@inproceedings{dinella2022toga,
author = {Dinella, Elizabeth and Ryan, Gabriel and Mytkowicz, Todd and Lahiri, Shuvendu K.},
title = {{TOGA}: a neural method for test oracle generation},
year = {2022},
isbn = {9781450392211},
publisher = {Association for Computing Machinery},
address = {New York, NY, USA},
url = {https://doi.org/10.1145/3510003.3510141},
doi = {10.1145/3510003.3510141},
booktitle = {Proceedings of the 44th International Conference on Software Engineering},
pages = {2130--2141},
numpages = {12},
location = {Pittsburgh, Pennsylvania},
series = {ICSE '22}
}

@inproceedings{zhao2026opsd,
  author = {Siyan Zhao and Zhihui Xie and Mengchen Liu and Jing Huang and Guan Pang and Feiyu Chen and Aditya Grover},
  title = {Self-Distilled Reasoner: On-Policy Self-Distillation for Large Language Models},
  booktitle = {Proceedings of the 43rd International Conference on Machine Learning},
  editor = {Tong Zhang and Miroslav Dudik and Martin Jaggi and Alekh Agarwal and Sharon Li and Dale Schuurmans and Jerry Zhu and Felix Berkenkamp and Hanze Dong and Alberto Bietti},
  series = {Proceedings of Machine Learning Research},
  volume = {306},
  pages = {162433--162448},
  year = {2026},
  month = jul,
  publisher = {PMLR},
  url = {https://proceedings.mlr.press/v306/zhao26be.html}
}

@inproceedings{widyasari2020bugsinpy,
  title     = {{BugsInPy}: A Database of Existing Bugs in {Python} Programs to Enable Controlled Testing and Debugging Studies},
  author    = {Ratnadira Widyasari and Sheng Qin Sim and Camellia Lok and
               Haodi Qi and Jack Phan and Qijin Tay and Constance Tan and
               Fiona Wee and Jodie Ethelda Tan and Yuheng Yieh and
               Brian Goh and Ferdian Thung and Hong Jin Kang and
               Thong Hoang and David Lo and Eng Lieh Ouh},
  booktitle = {Proceedings of the 28th ACM Joint Meeting on European Software
               Engineering Conference and Symposium on the Foundations of
               Software Engineering},
  pages     = {1556--1560},
  year      = {2020},
  doi       = {10.1145/3368089.3417943}
}

@article{guo2024deepseekcoder,
  title   = {{DeepSeek-Coder}: When the Large Language Model Meets Programming -- The Rise of Code Intelligence},
  author  = {Daya Guo and Qihao Zhu and Dejian Yang and Zhenda Xie and
             Kai Dong and Wentao Zhang and Guanting Chen and Xiao Bi and
             Yu Wu and Y. K. Li and Fuli Luo and Yingfei Xiong and
             Wenfeng Liang},
  journal = {arXiv preprint arXiv:2401.14196},
  year    = {2024}
}

\clearpage
\appendix

\section{Process Reward Model Details}
\label{app:prm_quality_analysis}

\subsection{PRM Data Quality Analysis}
\label{app:prm_data_quality}

We further analyze the quality of the automatically
constructed step-level labels used for PRM training.
As described in Section~\ref{sec:data_construction}, each candidate step is first
assigned an MC-derived label based on sampled continuations,
and is then verified by an LLM-based judge.
We retain a trajectory only if the MC-derived labels
and LLM judgments agree at every reasoning step,
and discard the entire trajectory otherwise.
The retained step-level labels are thus supported by both
execution-based evidence and semantic inspection.

\begin{table}[H]
\centering
\small
\setlength{\tabcolsep}{2.8pt}
\begin{tabular}{@{}lrrrrr@{}}
\toprule
Source & \#Traj. & \#Steps & \#Pos. & \#Neg. & Pos. \% \\
\midrule
MC labels & 55.2K & 165.7K & 92.6K &73.1K & 55.9 \\
LLM judgments & 55.2K & 165.7K &87.9K &77.8K & 53.0 \\
Consensus & 23.1K & 69.4K & 37.4K & 32.0K & 53.9 \\
\bottomrule
\end{tabular}
\caption{
Statistics of step-level labels used for PRM data construction.
\#Traj. denotes the number of response trajectories, and \#Steps denotes the number of step-level labels.
\#Pos. and \#Neg. denote the numbers of positive and negative step labels, respectively, while Pos. \% denotes the positive-label ratio.
The consensus set contains all step-level labels from
trajectories retained by this agreement criterion.
}
\label{tab:prm_data_quality}
\end{table}

Table~\ref{tab:prm_data_quality} reports the statistics of step-level labels used for PRM data construction.
Starting from 165.7K step labels annotated by both MC estimation and LLM judgment, we obtain 69.4K consensus labels from 23.1K response trajectories.
By requiring agreement between the two sources, the filtering step removes labels that are likely to be affected by either rollout noise in MC estimation or  subjective errors in LLM-based judgment.
The retained data contains 53.9\% positive steps and 46.1\% negative steps, suggesting that the filtering process improves label reliability without collapsing the supervision into a single dominant label.

\subsection{Process Reward Model Training}
\label{app:prm_training}

Bug-exposing test generation requires rigorous multi-step reasoning over program semantics, input design, and expected outputs. To effectively supervise this intermediate process, a Process Reward Model is essential.
However, existing PRMs are mostly trained on math or general reasoning~\cite{lightman2024lets,wang-etal-2024-math,zhu-etal-2025-retrieval}. To provide task-specific process supervision, we train a PRM on the step-level data from Section ~\ref{sec:data_construction}.

Concretely, we instantiate this as a binary step verification problem. Let $\mathcal{D}_{\mathrm{PRM}}$ denote the collection of all retained step instances. Each retained step instance is denoted as $e_t=(p,C_{\mathrm{buggy}},s_t,a_t,z_t)$, where $z_t \in \{0,1\}$ is the filtered step-level label. Let $q_t=(p,C_{\mathrm{buggy}},s_{t+1})$ denote the PRM input, where $s_{t+1}=\mathrm{Cat}(s_t,a_t)$. We train the PRM to predict a designated reward token $r(z_t) \in \{r^+,r^-\}$ at the end of the sequence. The training loss is defined as:
% \[
%     \mathcal{L}_{\mathrm{PRM}}
%     =
%     - \sum_{e_t \in \mathcal{D}_{\mathrm{PRM}}}
%     \log p_\phi\!\left(r(z_t)\mid q_t\right).
% \]
\begin{equation}
\mathcal{L}_{\mathrm{PRM}}
= -\sum_{e_t\in\mathcal{D}_{\mathrm{PRM}}}
\log p_\phi\!\left(r(z_t)\mid q_t\right).
\label{eq:prm_loss}
\end{equation}
After training, the predicted probability $p_\phi(r^+\mid q_t)$ is used as the process score of the current step.

\subsection{PRM Quality Verification}
\label{app:prm_quality_verification}

To further assess the reliability of the trained PRM, we construct an independent human-verified evaluation set that is completely excluded from PRM training.
The evaluation set contains 150 examples.
We manually inspect all step-level annotations and correct inconsistent labels according to the corresponding reasoning context.
Among the 450 labels, 404 are consistent with human judgments, while 46 are corrected, corresponding to a disagreement rate of 10.22\%.

We then evaluate the trained PRM on this human-corrected evaluation set.
For each step, the PRM predicts the probabilities of the positive and negative reward tokens, and the label with the higher probability is used as the binary prediction.
Table~\ref{tab:prm_quality_verification} reports the results.

\begin{table}[t]
\centering
\small
\setlength{\tabcolsep}{6pt}
\renewcommand{\arraystretch}{1.05}
\begin{tabular}{lcccc}
\toprule
\textbf{Model} &
\textbf{Accuracy} &
\textbf{Precision} &
\textbf{Recall} &
\textbf{F1} \\
\midrule
PRM & 77.11\% & 80.80\% & 78.60\% & 79.68\% \\
\bottomrule
\end{tabular}
\caption{PRM evaluation results on the independent human-verified step-level evaluation set.}
\label{tab:prm_quality_verification}
\end{table}
The PRM achieves 77.11\% accuracy, 80.80\% precision,
78.60\% recall, and 79.68\% F1.
These results indicate that the trained PRM maintains reasonable agreement with human judgments on unseen step-level reasoning instances.

\subsection{Effect of the MC Labeling Threshold}
\label{app:epsilon_analysis}

The relative-progress threshold $\epsilon$ controls how strictly
a step is labeled as positive during MC-based step labeling.
A smaller $\epsilon$ admits steps with lower relative MC scores,
which may introduce noisy positive labels, while a larger
$\epsilon$ imposes a stricter criterion and may exclude useful
intermediate steps.

We use PRM-guided beam search as a proxy evaluation of how
the labeling threshold affects the PRM's utility in test generation,
without repeatedly training the full test generator.
Specifically, we evaluate PRMs trained under different $\epsilon$
values on CodeContests+ using Qwen2.5-Coder-7B-Instruct
without task-specific post-training as the generator.
During search, the generator expands each retained reasoning
path with four candidate continuations.
The PRM scores all expanded paths, and we retain the three
highest-scoring paths for the next search step.
After the final search step, we select the highest-scoring
trajectory to produce a single test case for evaluation.
For the first reasoning step or a step whose preceding
prefix has an MC score of zero, we use an absolute
threshold of 0.5: the step is labeled positive if its
current MC score is at least 0.5, and negative otherwise.
The relative-progress threshold $\epsilon$ applies to
all remaining steps.

We report the joint success rate (PCR $\cap$ BTR), which
measures the percentage of evaluated instances for which
the generated test both triggers the bug and provides the
correct expected output.
Table~\ref{tab:epsilon_analysis} shows that the joint success
rate increases from 11.44\% to 15.10\% as $\epsilon$ rises
from 0.1 to 0.9, before decreasing to 13.96\% at $\epsilon=1.1$.
Under the $K=8$ MC estimator, 
$\epsilon=0.9$ accepts steps whose MC scores are at least
as high as those of their preceding prefixes, whereas
$\epsilon=1.1$ accepts only steps with higher MC scores.
The decrease at $\epsilon=1.1$ is consistent with a stricter
criterion excluding useful intermediate steps that preserve
the estimated success probability.
Based on this proxy evaluation, we select $\epsilon=0.9$
for our main experiments.

\begin{table}[H]
\centering
\small
\setlength{\tabcolsep}{3pt}
\renewcommand{\arraystretch}{1.05}
\begin{tabular*}{\columnwidth}{@{\extracolsep{\fill}}lcccccc@{}}
\toprule
\multicolumn{1}{c}{$\epsilon$}
& 0.1 & 0.5 & 0.7 & 0.8 & 0.9 & 1.1 \\
\midrule
PCR $\cap$ BTR (\%)
& 11.44 & 13.27 & 13.73 & 14.65 & 15.10 & 13.96 \\
\bottomrule
\end{tabular*}
\caption{
Effect of the MC labeling threshold $\epsilon$ on joint
success rate using PRM-guided beam search on CodeContests+.
The generator is Qwen2.5-Coder-7B-Instruct without
task-specific post-training.
}
\label{tab:epsilon_analysis}
\end{table}
\section{Process Reward Activation Timing}
\label{app:process_reward_activation}

In Section~\ref{sec:step_grpo}, we activate the PRM-based process reward only when the generated output is correct and the generated test case exposes the bug.
This section further analyzes this design choice by comparing different activation conditions for the process reward.

We conduct this analysis on the Qwen2.5-Coder-7B-Instruct backbone, using the same training data, PRM, and step-level GRPO configuration as the Step-GRPO baseline in our main experiments.
The comparison includes three variants:
\textit{Ungated PRM}, which applies the PRM-based process reward to all generated samples;
\textit{Subgoal-gated PRM}, which activates the process reward when either prediction correctness or bug triggering is satisfied;
and \textit{Full-success gated PRM}, which activates the process reward only when both objectives are satisfied and is the design adopted in \modelname.

\begin{table}[t]
\centering
\small
\setlength{\tabcolsep}{6pt}
\begin{tabular}{lcc}
\toprule
\textbf{Activation Strategy} & \textbf{PCR} & \textbf{PCR$\,\cap\,$BTR} \\
\midrule
Ungated PRM
& 78.26\% & 23.34\% \\

Subgoal-gated PRM
& 80.32\% & 24.03\% \\

\rowcolor{gray!15}
Full-success gated PRM
& \textbf{84.21\%} & \textbf{27.69\%} \\
\bottomrule
\end{tabular}
\caption{
Effect of different activation conditions for the PRM-based process reward on CodeContests+.
All variants use the same backbone, training data, PRM, and RL algorithm, and differ only in when $R_{\mathrm{proc}}$ is activated.
}
\label{tab:process_reward_activation}
\end{table}

Table~\ref{tab:process_reward_activation} shows that applying the PRM-based process reward more broadly does not necessarily improve final task performance.
The ungated variant obtains the lowest joint success rate of 23.34\%.
Activating the PRM when either sub-objective is satisfied increases the joint success rate by 0.69 percentage points to 24.03\%.
The full-success gated strategy achieves the best performance on both metrics, with a PCR of 84.21\% and a joint success rate of 27.69\%.
Its joint success rate exceeds those of the ungated and subgoal-gated variants by 4.35 and 3.66 percentage points, respectively.

These results suggest that the PRM should not be used as an unconditional reward.
When PRM scores are assigned to failed or partially successful trajectories, the policy may optimize toward patterns favored by the PRM rather than toward the final task objective.
This can make the training process vulnerable to reward hacking, especially because the PRM is trained to evaluate meaningful intermediate reasoning steps but may behave unpredictably on out-of-distribution responses.

\newcommand{\codefence}{\texttt{\textasciigrave\textasciigrave\textasciigrave}}

\begin{figure}[t]
\small
\begin{tcolorbox}[
  colback=gray!5,
  colframe=gray!50,
  title=Reward-hacking example from Ungated PRM,
  fonttitle=\bfseries,
]
\textbf{\#\#\# Bug Analysis}\\
Logical flaw.\\[4pt]

\textbf{\#\#\# Test Design}\\[-2pt]

\textbf{\#\#\# Output Prediction}\\[-2pt]

\textbf{\#\#\# Input}

\codefence

\begin{verbatim}
0
\end{verbatim}

\codefence

\textbf{\#\#\# Expected Output}

\codefence

\begin{verbatim}
a
\end{verbatim}

\codefence

\end{tcolorbox}
\caption{
A reward-hacking example produced by the ungated PRM variant.
The model keeps the required response structure but provides almost no meaningful reasoning process.
}
\label{fig:reward_hacking_example}
\end{figure}

Figure~\ref{fig:reward_hacking_example} illustrates this failure mode.
The response preserves the required section headers, but the intermediate reasoning is either empty or semantically uninformative.
Such responses are outside the intended input distribution of the PRM, which is designed to judge meaningful reasoning steps rather than empty process templates.
When the PRM is applied to all samples, the policy can exploit this weakness by producing degenerate reasoning formats that may receive favorable process scores without actually supporting the generated test case.

Therefore, \modelname adopts the full-success gated strategy.
By activating process rewards only when both prediction correctness and bug triggering are satisfied, this strategy anchors PRM-based optimization to execution-verified task success and reduces incentives to exploit process scores on failed trajectories.
The PRM remains useful after gating because successful trajectories can still differ in how they arrive at a valid test: some rely on accurate bug analysis and targeted test design, while others succeed despite flawed or redundant reasoning.
Process rewards distinguish these reasoning processes and encourage the policy to reinforce higher-quality steps within successful trajectories~\citep{Posterior-GRPO}.
Thus, full-success gating provides informative process supervision while  mitigating reward hacking.

\section{Training Curve Analysis}
\label{sec:training_dynamics}

\begin{figure*}[t]
    \centering
    \includegraphics[width=\textwidth]{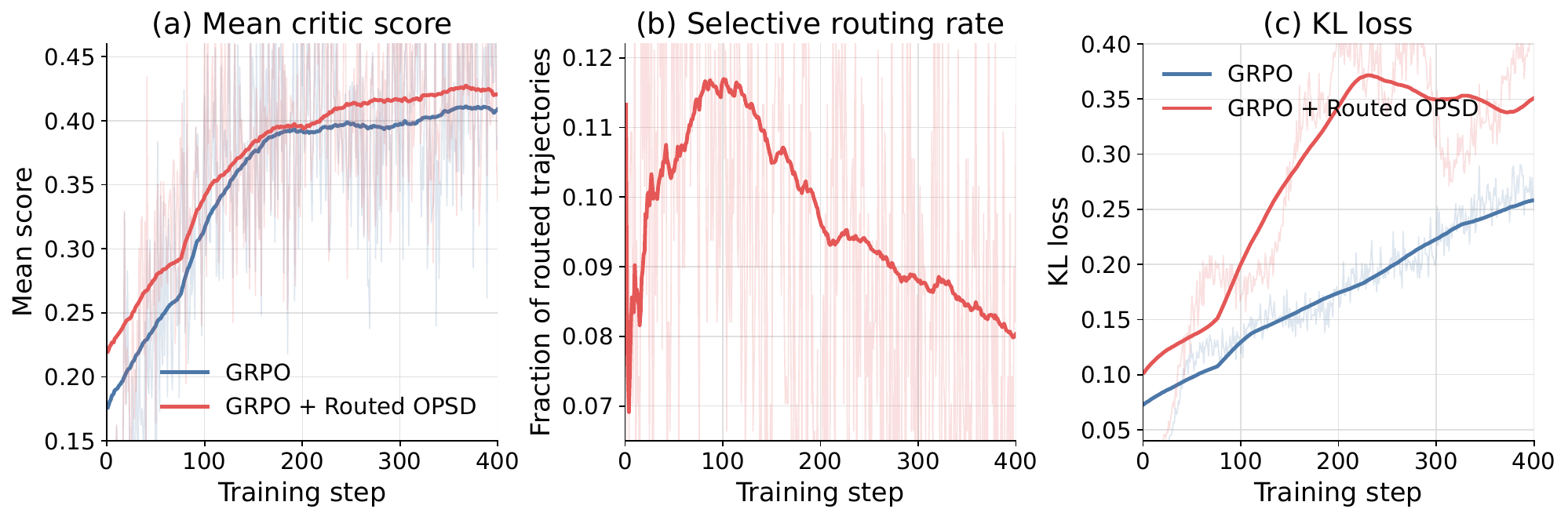}
    \caption{
    Training dynamics of standard GRPO and GRPO with failure-routed OPSD.
    (a) Mean score during training.
    (b) Fraction of trajectories routed to OPSD.
    (c) KL loss during training.
    }
    \label{fig:training_dynamics}
\end{figure*}

To further understand how failure-routed OPSD affects the optimization process, we compare the training dynamics of standard GRPO and GRPO with routed OPSD.
Figure~\ref{fig:training_dynamics} reports three complementary statistics throughout training: the mean training score, the fraction of trajectories routed to OPSD, and the KL loss.
Together, these curves characterize both the optimization progress of the policy and the behavior of the selective routing mechanism.

\textbf{Selective routing is sufficient to produce stronger policy improvement.}
As shown in Figure~\ref{fig:training_dynamics}(a), GRPO with routed OPSD achieves a higher mean score than standard GRPO after the early training stage and converges to a better final level.
Figure~\ref{fig:training_dynamics}(b) shows that only about 8--12\% of trajectories are routed to OPSD, indicating that selectively recovering a small subset of partial-success trajectories is sufficient to provide useful additional supervision.
The routing rate further decreases in later training, suggesting that fewer trajectories require targeted recovery as the policy improves.

\textbf{Failure-routed OPSD induces larger but meaningful policy updates.}
Figure~\ref{fig:training_dynamics}(c) shows that routed OPSD produces a consistently larger KL divergence than standard GRPO, indicating a stronger policy shift.
Importantly, this larger deviation is accompanied by the higher mean score in Figure~\ref{fig:training_dynamics}(a), suggesting that the additional movement reflects task-relevant improvement rather than unstable drift.
Meanwhile, the KL curve gradually stabilizes, indicating that these stronger updates eventually reach a stable regime.
\section{Additional Generalization Results}
\label{app:generalization}

We further evaluate the generalization of \modelname from two
perspectives: generalization to a different model family and
generalization to real-world software bugs.
These experiments examine whether the effectiveness of \modelname
depends on the Qwen2.5-Coder backbone or on the
competitive-programming-style data used for training.

\subsection{Generalization to Another Model Family}
\label{app:cross_model}

Our main experiments use Qwen2.5-Coder as the generator backbone.
To examine whether \modelname transfers to a different model family,
we additionally conduct experiments with
DeepSeek-Coder-6.7B-Instruct (abbreviated as DSC-6.7B) \citep{guo2024deepseekcoder}.
We keep the training data, reward definitions, routing criterion, and
evaluation protocol unchanged, and replace only the generator backbone.
The task-specific PRM is kept fixed so that the experiment isolates the
effect of changing the policy model family.

Table~\ref{tab:cross_model_generalization} reports the results on
CodeContests+.
Compared with standard GRPO, \modelname improves PCR from
59.27\% to 64.07\%, BTR from 27.92\% to 29.52\%, and
PCR$\,\cap\,$BTR from 18.76\% to 20.37\%, corresponding to gains of
4.80, 1.60, and 1.61 percentage points, respectively.
These results show that the improvements brought by step-level credit
assignment and failure-routed OPSD are not specific to the
Qwen2.5-Coder generator backbone.

\begin{table}[t]
\centering
\small
\setlength{\tabcolsep}{5pt}
\renewcommand{\arraystretch}{1.08}
\begin{tabular}{lccc}
\toprule
\textbf{Method}
& \textbf{PCR}
& \textbf{BTR}
& \textbf{PCR$\,\cap\,$BTR} \\
\midrule
DSC-6.7B
& 27.69\%
& 16.93\%
& 8.01\% \\
 GRPO
& 59.27\%
& 27.92\%
& 18.76\% \\
\textbf{\modelname}
& \textbf{64.07\%}
& \textbf{29.52\%}
& \textbf{20.37\%} \\
\bottomrule
\end{tabular}
\caption{
Cross-family generalization results on CodeContests+ using
DSC-6.7B as the generator backbone.
}
\label{tab:cross_model_generalization}
\end{table}

\subsection{Generalization to 
Real-World Software Bugs}
\label{app:real_world_generalization}

The benchmarks used in our main experiments mainly consist of
programming problems paired with incorrect solutions.
To further examine whether \modelname transfers to real-world software
bugs, we construct an additional evaluation set from BugsInPy \citep{widyasari2020bugsinpy}, which
contains real bugs collected from open-source Python projects.
We retain 30 instances whose relevant code context fits within the
model context window.
For each instance, we provide the buggy code together with the
dependency context required to understand and execute the target
behavior, thereby avoiding repository-level retrieval as an additional
confounding factor.
No BugsInPy instances are used during \modelname training.

We use Qwen2.5-Coder-7B-Instruct (abbreviated as QWC-7B) as the
generator backbone.
Table~\ref{tab:real_world_generalization} reports the results.
Compared with standard GRPO, \modelname improves PCR from
60.00\% to 66.67\%, BTR from 33.33\% to 40.00\%, and
PCR$\,\cap\,$BTR from 20.00\% to 26.67\%, corresponding to gains of
6.67 percentage points on all three metrics.
These results indicate that the effectiveness of \modelname extends
beyond the training-domain benchmarks to real-world software bugs.

\begin{table}[t]
\centering
\small
\setlength{\tabcolsep}{5pt}
\renewcommand{\arraystretch}{1.08}
\begin{tabular}{lccc}
\toprule
\textbf{Method}
& \textbf{PCR}
& \textbf{BTR}
& \textbf{PCR$\,\cap\,$BTR} \\
\midrule
QWC-7B
& 26.67\%
& 23.33\%
& 6.67\% \\
 Standard GRPO
& 60.00\%
& 33.33\%
& 20.00\% \\
\textbf{\modelname}
& \textbf{66.67\%}
& \textbf{40.00\%}
& \textbf{26.67\%} \\
\bottomrule
\end{tabular}
\caption{
Out-of-domain generalization results on the real-world
BugsInPy evaluation set.
}
\label{tab:real_world_generalization}
\end{table}

Overall, these additional experiments provide complementary evidence
that \modelname generalizes across both generator model families and task
domains, rather than relying on a particular backbone or benchmark
distribution.

\flushbottom
\section{More Repair Results}
\label{app:repair_more_results}

We provide additional program repair results under different numbers of repair attempts.
In Section~\ref{sec:application}, we report the main results with $K_2=3$.
Here, we further evaluate $K_2=1$ and $K_2=5$ while varying the number of sampled generated tests $K_1$.

\begin{figure}[t]
    \centering
    \includegraphics[width=\columnwidth]{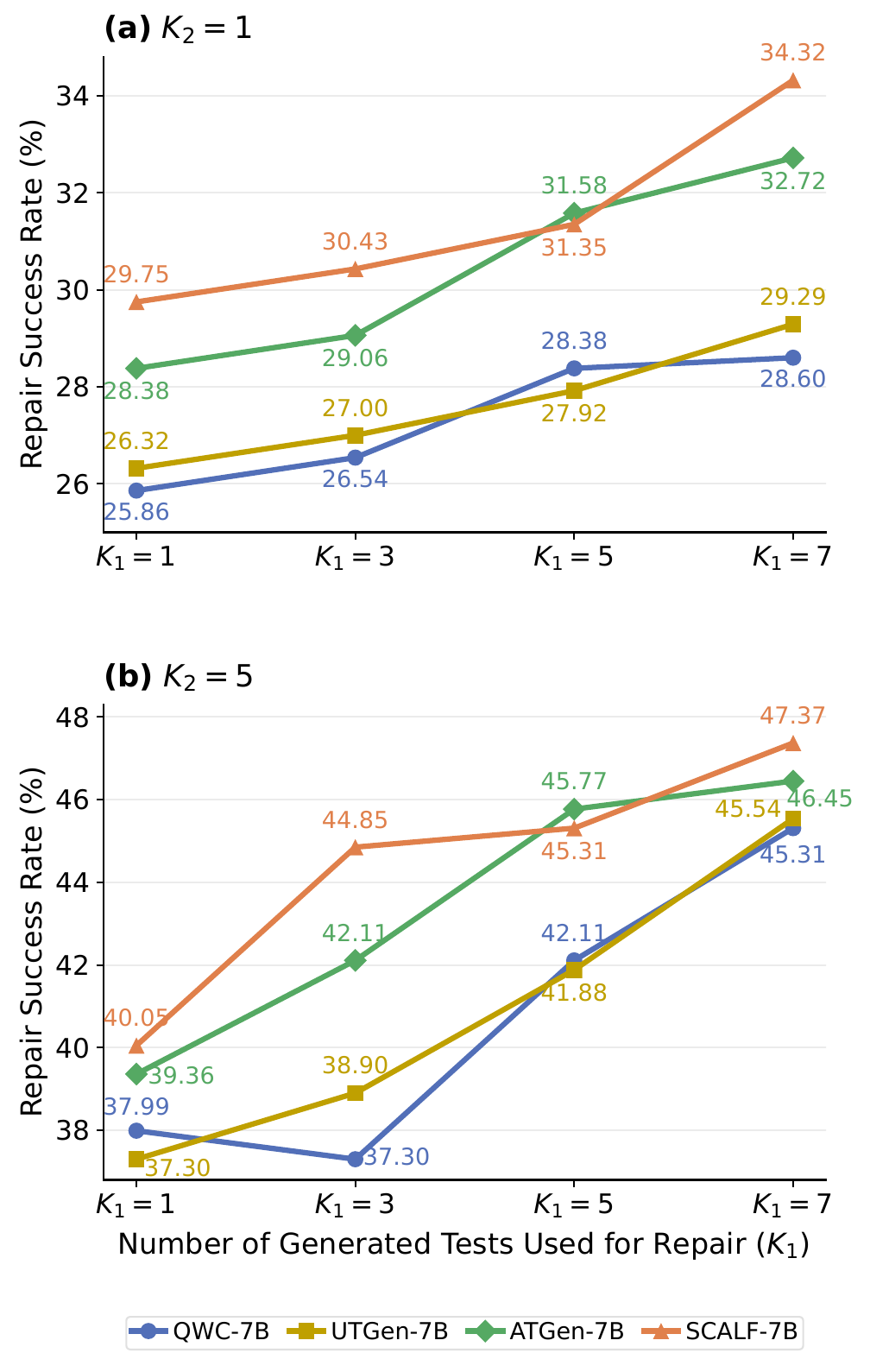}
    \caption{
    Additional program repair results under different numbers of repair attempts.
    (a) Results with $K_2=1$.
    (b) Results with $K_2=5$.
    }
    \label{fig:repair_k2_1_5}
\end{figure}

Figure~\ref{fig:repair_k2_1_5} shows that \modelname-7B remains effective under both smaller and larger numbers of repair attempts.
When $K_2=1$, \modelname-7B achieves the best repair success rate at three of the four $K_1$ values and reaches 34.32\% at $K_1=7$, indicating that its generated tests can provide useful failure evidence even when the repair model has only one repair attempt per test.
When $K_2=5$, all methods benefit from more repair attempts, while \modelname-7B achieves the best result at $K_1=7$ with 47.37\%.
These results further support the conclusion that \modelname-7B generates effective test cases for test-feedback-driven program repair across different repair settings.

\section{Implementation Details}
\label{app:implementation_details}

\subsection{Training Configuration}
\label{app:training_configuration}

Table~\ref{tab:prm_training_details} and
Table~\ref{tab:generator_training_details} summarize the key training
hyperparameters used in our experiments.
UTGen and ATGen share the applicable common training settings
with \modelname, while method-specific hyperparameters apply
only to the corresponding methods.
All experiments are conducted on 8 NVIDIA H100 GPUs.
A single model training run takes approximately 26 hours on 8 NVIDIA H100 GPUs, corresponding to approximately 208 GPU-hours per run.
For evaluation, all models are decoded with a temperature of 0.8.
\begin{table}[t]
\centering
\small
\setlength{\tabcolsep}{6pt}
\renewcommand{\arraystretch}{1.05}
\begin{tabular}{p{0.48\columnwidth}p{0.38\columnwidth}}
\toprule
\textbf{Hyperparameter} & \textbf{Value} \\
\midrule
Training framework & OpenRLHF \\
Training epochs & 1 \\
Learning rate & $1\times10^{-6}$ \\
Global batch size & 32 \\
Micro-batch size & 2 \\
Maximum sequence length & 4096 \\
Precision & bf16 \\
ZeRO stage & 2 \\
\bottomrule
\end{tabular}
\caption{Hyperparameters for PRM training.}
\label{tab:prm_training_details}
\end{table}

\begin{table}[t]
\centering
\small
\setlength{\tabcolsep}{6pt}
\renewcommand{\arraystretch}{1.05}
\begin{tabular}{p{0.48\columnwidth}p{0.38\columnwidth}}
\toprule
\textbf{Hyperparameter} & \textbf{Value} \\
\midrule
Training framework & veRL \\
Training epochs & 2 \\
Learning rate & $1\times10^{-6}$ \\
Learning-rate scheduler & Cosine \\
Training batch size & 32 \\
PPO mini-batch size & 32 \\
PPO micro-batch size per device & 1 \\
PPO clipping parameter ($\epsilon_c$) & 0.2 \\
Reference KL coefficient ($\beta$) & 0.001 \\
Maximum prompt length & 2048 \\
Maximum response length & 2048 \\
Rollout engine & vLLM \\
Rollouts per prompt & 8 \\
Rollout temperature & 0.7 \\
Rollout top-p & 0.95 \\
Tensor parallel size & 2 \\
Entropy coefficient & 0 \\
OPSD loss weight ($\lambda_{\mathrm{OPSD}}$) & 0.1 \\
OPSD pointwise clipping threshold & 0.05 \\
Random seed & 42 \\
\bottomrule
\end{tabular}
\caption{Hyperparameters for Test Generator training.}
\label{tab:generator_training_details}
\end{table}

\subsection{Training Algorithm}
\label{app:training_algorithm}

Algorithm~\ref{alg:scalf_training} summarizes the training procedure
of \modelname described in Section~\ref{sec:method}.
For each task, the policy samples a group of trajectories and obtains
execution feedback for prediction correctness and bug triggering.
PRM scores are added to the outcome reward only for trajectories
satisfying both objectives.
Step-level advantages are then computed across the entire rollout
group, including trajectories subsequently routed to self-distillation.

The execution feedback also determines the optimization route.
Trajectories that trigger the bug but predict an incorrect output
are optimized through OPSD: the current policy uses reference-derived
feedback to form a teacher distribution and supervises its
unconditioned student distribution on the same sampled prefixes.
All remaining trajectories are optimized through Step-GRPO,
with each step's advantage applied to its corresponding token span.
The losses from both routes are aggregated to update the policy.

\begin{algorithm*}[t]
\caption{Training Procedure of \modelname}
\label{alg:scalf_training}
\small
\begin{algorithmic}[1]
\Require Training tasks
$\mathcal{D}=\{(p,C_{\mathrm{buggy}},C_{\mathrm{ref}})\}$;
initial policy $\pi_\theta$
\Require Fixed PRM $p_\phi$; fixed reference policy $\pi_{\mathrm{ref}}$
\Require Training epochs $E=2$; rollout group size $G=8$;
clipping parameter $\epsilon_c=0.2$
\Require OPSD loss weight $\lambda_{\mathrm{OPSD}}=0.1$;
KL coefficient $\beta$; stability constant $\delta>0$
\Ensure Trained policy $\pi_\theta$

\For{$e=1,\ldots,E$}
    \For{each training mini-batch $\mathcal{B}$ from $\mathcal{D}$}
        \State $\theta_{\mathrm{old}}\gets\theta$

        \For{each $(p,C_{\mathrm{buggy}},C_{\mathrm{ref}})\in\mathcal{B}$}
            \State $c\gets(p,C_{\mathrm{buggy}})$
            \State Sample
            $\{\tau_i\}_{i=1}^{G}\sim
            \pi_{\theta_{\mathrm{old}}}(\cdot\mid c)$

            \For{$i=1,\ldots,G$}
                \State Extract test $(x_i,y_i)$ and
                reasoning-step token spans $\{S_{i,t}\}_{t=1}^{T}$
                \State Evaluate format and test execution to obtain
                $R_i^{\mathrm{fmt}}$, $BT_i$, $PC_i$,
                and execution feedback
                \State
                $R_i^{\mathrm{out}}\gets
                BT_i+PC_i+R_i^{\mathrm{fmt}}$

                \If{$BT_i=1$ and $PC_i=1$}
                    \State Score each reasoning step with the fixed
                    PRM to obtain $\{p_{i,t}\}_{t=1}^{T}$
                    \State
                    $\widetilde{R}_{i,t}\gets
                    R_i^{\mathrm{out}}+p_{i,t}$
                    for all $t$
                \Else
                    \State
                    $\widetilde{R}_{i,t}\gets R_i^{\mathrm{out}}$
                    for all $t$
                \EndIf
            \EndFor

            \For{$t=1,\ldots,T$}
                \State
                $\mu_t\gets
                \frac{1}{G}\sum_{i=1}^{G}\widetilde{R}_{i,t}$
                \Comment{Include all trajectories}
                \State
                $\sigma_t\gets
                \operatorname{Std}
                \big(\{\widetilde{R}_{i,t}\}_{i=1}^{G}\big)$
                \State
                $A_{i,t}\gets
                (\widetilde{R}_{i,t}-\mu_t)/(\sigma_t+\delta)$
                for all $i$
            \EndFor

            \For{$i=1,\ldots,G$}
                \State
                $z_i\gets
                \mathbb{I}[BT_i=1\land PC_i=0]$

                \If{$z_i=1$}
                    \State Retrieve the reference output
                    $y_i^{*}=C_{\mathrm{ref}}(x_i)$
                    \State Construct privileged information $f_i$
                    from $(x_i,y_i^{*})$ and execution feedback
                    \State
                    $\pi^{\mathrm{S}}_{i,k}\gets
                    \pi_\theta(\cdot\mid c,u_{i,<k})$
                    for all tokens $k$
                    \State
    $\pi^{\mathrm{T}}_{i,k}\gets
    \operatorname{stopgrad}\!\left[
    \pi_\theta(\cdot\mid c,u_{i,<k},f_i)
    \right]$
    for all tokens $k$
    \Comment{No gradient through the teacher}
                    \State
                    $\mathcal{L}_{\mathrm{OPSD},i}\gets
                    \frac{1}{|\tau_i|}
                    \sum_{k=1}^{|\tau_i|}
                    D_{\mathrm{KL}}
                    \big(\pi^{\mathrm{S}}_{i,k}
                    \,\|\,\pi^{\mathrm{T}}_{i,k}\big)$
                    \State
                    $\ell_i\gets
                    \lambda_{\mathrm{OPSD}}
                    \mathcal{L}_{\mathrm{OPSD},i}$
                \Else
                    \State Compute token-wise policy ratios
                    $\rho_{i,k}(\theta)$ against
                    $\pi_{\theta_{\mathrm{old}}}$
                    \State Apply $A_{i,t}$ only to tokens
                    $k\in S_{i,t}$
                    \State Compute
                    $\mathcal{L}_{\mathrm{step\text{-}GRPO},i}$
                    using clipping parameter $\epsilon_c$ and
                    KL regularization coefficient $\beta$,
                    as defined in Section~\ref{sec:step_grpo}
                    \State
                    $\ell_i\gets
                    \mathcal{L}_{\mathrm{step\text{-}GRPO},i}$
                \EndIf
            \EndFor

            \State
            $\mathcal{L}_c\gets
            \frac{1}{G}\sum_{i=1}^{G}\ell_i$
        \EndFor

        \State Update $\theta$ by minimizing
        $\frac{1}{|\mathcal{B}|}
        \sum_{c\in\mathcal{B}}\mathcal{L}_c$
        with the training optimizer
    \EndFor
\EndFor

\State \Return $\pi_\theta$
\end{algorithmic}
\end{algorithm*}

\section{Prompt Template}
\label{app:prompt}

\begin{figure*}[!t]
\centering
\small
\begin{tcolorbox}[
  width=\textwidth,
  colback=gray!5,
  colframe=gray!50,
  title=The Prompt for Test Case Generation,
  fonttitle=\bfseries,
]
You are given a programming task description and an incorrect code solution. Your goal is to identify the bug in the code and construct a Hack Test Case that triggers the failure of the solution.

\textbf{\#\# Task Description}

\textcolor{promptblue}{\{description\}}

\textbf{\#\# Incorrect Code Solution}

\codefence

\textcolor{promptblue}{\{code\}}

\codefence

\textbf{\#\# Response Format}

Your response MUST exactly follow the format below:

\textbf{\#\#\# Bug Analysis}

Briefly describe the logic or implementation error in the code.

\textbf{\#\#\# Test Design}

Based on the bug analysis, describe the input values chosen to trigger the bug.

\textbf{\#\#\# Output Prediction}

Concisely derive the correct output based on the Task Description logic.

\textbf{\#\#\# Input}

Provide the raw standard input that follows the exact input specification. Do NOT include ANY reasoning, explanations, or markdown formatting.

\textbf{\#\#\# Expected Output}

Provide the correct standard output according to the task description. Do NOT include ANY reasoning, explanations, or markdown formatting.
\end{tcolorbox}
\caption{Prompt template for test-case generation.}
\label{fig:prompt}
\end{figure*}

\begin{figure*}[!t]
\centering
\small
\begin{tcolorbox}[
  width=\textwidth,
  colback=gray!5,
  colframe=gray!50,
  title=The Prompt for Program Repair,
  fonttitle=\bfseries,
]
You are given a programming task description and an \textbf{incorrect} code solution.
Your goal is to repair the incorrect code according to the task specification.
A failing test case is provided as evidence: it includes the test input, the correct output, and the output produced by the incorrect code.
Use this evidence to identify the bug and then repair the code.

\textbf{\#\# Task Description}

\textcolor{promptblue}{\{description\}}

\textbf{\#\# Incorrect Code Solution}

\textcolor{promptblue}{\{code\}}

\textbf{\#\# Failing Test Input}

\textcolor{promptblue}{\{test\_input\}}

\textbf{\#\# Correct Output}

\textcolor{promptblue}{\{expected\_output\}}

\textbf{\#\# Incorrect Code Output}

\textcolor{promptblue}{\{buggy\_output\}}

\textbf{\#\# Important Notes}

The repaired code must solve the original task generally and preserve the required input/output behavior.

\textbf{\#\# Response Format}

Think step by step.

Finally, output only the complete repaired code wrapped in a single code block.

Do not output any explanation outside the code block.

\textcolor{promptblue}{\texttt{\textless complete repaired code\textgreater}}
\end{tcolorbox}
\caption{Prompt template for program repair.}
\label{fig:repair_prompt}
\end{figure*}
In this section, we present the prompts used in training and inference for an LLM to perform various operations.

Figures~\ref{fig:prompt} and~\ref{fig:repair_prompt} present the prompt templates used for test-case generation and program repair, respectively.

\end{document}